\documentclass[onecolumn,noshowpacs,nofootinbib,11pt]{revtex4-1}
\usepackage{array}
\newcommand*\xoverline[2][0.75]{%
    \sbox{\myboxA}{$\m@th#2$}%
    \setbox\myboxB\null
    \ht\myboxB=\ht\myboxA%
    \dp\myboxB=\dp\myboxA%
    \wd\myboxB=#1\wd\myboxA
    \sbox\myboxB{$\m@th\overline{\copy\myboxB}$}
    \setlength\mylenA{\the\wd\myboxA}
    \addtolength\mylenA{-\the\wd\myboxB}%
    \ifdim\wd\myboxB<\wd\myboxA%
       \rlap{\hskip 0.5\mylenA\usebox\myboxB}{\usebox\myboxA}%
    \else
        \hskip -0.5\mylenA\rlap{\usebox\myboxA}{\hskip 0.5\mylenA\usebox\myboxB}%
    \fi}
\makeatother
\makeatletter
\newsavebox\myboxA
\newsavebox\myboxB
\newlength\mylenA
\usepackage{float,xcolor,upgreek,yfonts,tikz}
\usepackage{color}
\usepackage{bm}
\usepackage{comment}
\usepackage{graphicx,bigints}
\usepackage{graphicx,float}\usepackage[all]{xy}
\usepackage{amsmath,amssymb,upgreek}

   \usepackage{caption}
   \newcommand{\clt}{\textcolor{black}}
 \makeatletter
\newcommand{\thickhline}{%
    \noalign {\ifnum 0=`}\fi \hrule height 1pt
    \futurelet \reserved@a \@xhline
}
\newcolumntype{"}{@{\hskip\tabcolsep\vrule width 1pt\hskip\tabcolsep}}
\makeatother\usepackage{booktabs,makecell}
\usepackage{subcaption}
\usepackage{physics}
\usepackage{tensor}
\usepackage{amssymb}

\usepackage{alphabeta}
 \usepackage{cancel}
\usepackage{dcolumn}
\usepackage{textgreek}
\usepackage{adjustbox}
\usepackage{multirow} 
\usepackage{amsmath}
\allowdisplaybreaks
\usepackage{units}
\usepackage{overpic}

\newcommand{\beq}{\begin{eqnarray}}
\newcommand{\eeq}{\end{eqnarray}}
\newcommand{\be}{\begin{equation}}
\newcommand{\ee}{\end{equation}}

\newcommand{\bea}{\begin{eqnarray}}
\newcommand{\eea}{\end{eqnarray}}

\newcommand{\ba}{\begin{eqnarray}}
\newcommand{\ea}{\end{eqnarray}}

\newcommand{\cvm}{current version of the manuscript\;}

\newcommand{\pvm}{previously submitted version of the manuscript\;}

\usepackage[colorlinks,hyperindex,unicode]{hyperref}
\definecolor{green1}{RGB}{0,128,0} 
\hypersetup{hidelinks,backref=true,pagebackref=true,hyperindex=true,colorlinks=true,breaklinks=true,urlcolor= blue}
\hypersetup{%
  colorlinks = true,
  linkcolor  = blue,
  citecolor = cyan,
}
\usepackage{bookmark,textgreek}
\usepackage{hyperref,color,xcolor}
\hypersetup{hidelinks,hyperindex=true,colorlinks=true,breaklinks=true,urlcolor= blue}
\hypersetup{%
  colorlinks = true,
  linkcolor  = blue
}

\newcommand\orcidguimaraes{{\href{https://orcid.org/0009-0002-6897-1412}{\orcidicon}}}

\newcommand\orcidrogerio{{\href{https://orcid.org/0000-0001-7848-5472}{\orcidicon}}}

\newcommand\orcidroldao{{\href{https://orcid.org/0000-0003-3978-532X}{\orcidicon}}}
\newcommand{\orcidicon}{%
	\begin{tikzpicture}
	\draw[lime, fill=lime] (0,0)
		circle [radius=0.16]
		node[white] {{\fontfamily{qag}\selectfont \tiny ID}};
	\draw[white, fill=white] (-0.0625,0.095)
		circle [radius=0.007];
	\end{tikzpicture}	\hspace{-2mm}
}

\begin{document}
\title{
Hair imprints of the gravitational decoupling and hairy black hole spectroscopy}

\author{V. F. Guimar\~aes\orcidguimaraes}
\affiliation{Center for Natural and Human Sciences, Federal University of ABC, 09210-580, Santo Andr\'e, Brazil}
\email{vitor.guimaraes@ufabc.edu.br}

\author{R. T. Cavalcanti\orcidrogerio}
\affiliation{Institute of Mathematics and Statistics, Rio de Janeiro State University, 20550-900, Rio de Janeiro, Brazil}
\affiliation{Physics department, São Paulo State University, 12516-410, Guaratingueta, Brazil}
\email{rogerio.cavalcanti@ime.uerj.br}

\author{R. da Rocha\orcidroldao}
\affiliation{Center of Mathematics, Federal University of ABC, 09210-580, Santo Andr\'e, Brazil}
\email{roldao.rocha@ufabc.edu.br (corresponding author)}

\medbreak
\begin{abstract} 
Hairy black holes by gravitational decoupling (GD) are probed to derive the gravitational waveform produced by perturbation theory applied to these compact objects. Using the Regge--Wheeler and Zerilli equations governing the metric perturbations and applying a higher-order WKB method, the quasinormal modes (QNMs) are computed and discussed. Compared to the QNMs produced in the ringdown phase of Reissner--Nordström black hole solutions, it yields a clear physical signature of primary hair imprinting the hairy GD black hole gravitational waveforms.
\end{abstract}



\maketitle

\section{Introduction}
Laser-interferometric detection of gravitational waves (GWs) radiated from the final stages
of merging binaries has contributed to understanding the strong regime of gravity \cite{LIGOScientific:2016lio,LIGOScientific:2019fpa}.
This remarkable achievement allows for experimental validation of solutions of the Einstein field equations and their extensions. The gravitational decoupling (GD) technique is a useful and powerful method to derive self-gravitating compact stellar configurations from established solutions in general relativity (GR). This approach naturally leads to the emergence of anisotropic stellar distributions, facilitating the derivation of analytical solutions to the Einstein field equations by incorporating diverse possibilities for the energy-momentum tensor~\cite{Ovalle:2017fgl,Ovalle:2018gic}.
Within the GD framework, the sources of the GR gravitational field and their corresponding field equations are meticulously split into two disjoint sectors. The first describes the standard GR solution, while the second encompasses additional sources that can embody various forms of charge, ranging from tidal and gauge charges to hairy fields, along with contributions from extended gravity models. A plethora of physically relevant compact stellar configurations derived from the GD method can be found in Refs. ~\cite{Estrada:2019aeh, Gabbanelli:2019txr, Leon:2023nbj,Ramos:2021drk, Sharif:2023ecm, Rincon:2019jal, Morales:2018urp, Panotopoulos:2018law, Singh:2019ktp, Jasim:2023ehu}. Realistic models based on the relativistic description of nuclear interactions also indicate that the interior of stars exhibits anisotropies at extremely high densities. The GD method naturally accommodates pressure anisotropies~\cite{Gabbanelli:2018bhs, PerezGraterol:2018eut, Heras:2018cpz, Torres:2019mee, Hensh:2019rtb, Contreras:2019iwm, Tello-Ortiz:2021kxg, Andrade:2023wux, Zubair:2023cvu, Bamba:2023wok, Maurya:2023uiy, Iqbal:2025xqf, Tello-Ortiz:2023poi, Contreras:2021xkf, Sharif:2020lbt}. The GD method can be approached within the gauge/gravity duality, allowing for a comprehensive exploration of physically viable black holes in the infrared limit \cite{Meert:2020sqv,DaRocha:2019fjr,daRocha:2020gee,Estrada:2024lhk,daRocha:2014dla}. Additionally, the GD method has been explored and tested by strong deflection limit lensing effects \cite{Cavalcanti:2016mbe}. 
Ref. \cite{Ovalle:2020kpd} introduced the prominent class of hairy GD black holes, which has been broadly applied to relevant theoretical developments \cite{Liang:2024xif,Avalos:2023ywb,Zhang:2022niv,Ditta:2023arv,Mansour:2024mdg,Mahapatra:2022xea, Albalahi:2024vpy,  Naseer:2024jjy, Maurya:2024zao, Misyura:2024fho}. 
Recent studies have also addressed observational aspects of GD hairy black holes. In this context, quasinormal modes (QNMs)  emitted from GD hairy  solutions, including other important observational aspects, were scrutinized in Refs. \cite{Cavalcanti:2022cga,Yang:2022ifo,Li:2022hkq,Cavalcanti:2022adb,Priyadarshinee:2023exb,Rehman:2023eor,Avalos:2023jeh,Al-Badawi:2024iax,Tello-Ortiz:2024mqg}.

Hairy black holes exhibit additional macroscopic degrees of freedom unrelated to quasilocal conserved quantities at the event horizon. How the microscopic description of hairy GD black holes accounts for these extra degrees of freedom, representing primary hair, suggests an effective approach for deriving analytical solutions. With the first unambiguous observations of GWs by LIGO/Virgo, which directly capture perturbations in the curvature of the spacetime weaving, hairy GD solutions can be thoroughly examined in GW  astrophysics. It supports exploring aspects of gravity in the strongly nonlinear regime and assessing any deviations from the predictions of GR. 
Coalescing binary black hole systems, particularly during the merger ringdown phase, have been thoroughly analyzed from a direct observational perspective, yielding significant results \cite{LIGOScientific:2018dkp} that may directly  provide new insights to hairy GD  solutions.

Numerical relativity is an essential tool within the fully nonlinear regime of GR, supporting significant results that also stem from the post-Newtonian regime and black hole perturbation theory. Despite substantial progress, there remain unanswered questions regarding the foundations of GR, such as, e.g., the lack of a complete quantum field theory of gravity and the physical interpretation of gravitational singularities. The range of extensions to GR regarding black hole solutions, associated with the increased sensitivity of future gravitational-wave detectors, may provide an unprecedented opportunity to test the foundational aspects of gravity \cite{will2014confrontation}, particularly the hairy GD solutions.

Since the pioneering work of Regge and Wheeler \cite{Regge:1957td}, black hole perturbation theory has played a crucial role in exploring the strong field regime of GR \cite{Cardoso:2019rvt,BarackCardosoNissankeSotiriou,Richarte:2021fbi}. With the increasing interest in GWs, black hole perturbation theory and its various applications have become essential tools in gravitational physics  \cite{Pani:2013,BarackCardosoNissankeSotiriou}. A key setup involves investigating QNMs, which characterize a black hole relaxation process after external perturbations, with several relevant developments in Refs. \cite{Miranda:2009uw,Lin:2025zea,Anacleto:2021qoe,Oliveira:2018oha,Rougemont:2018ivt}.  Perturbation theory also successfully describes binary black hole mergers, during which black holes emit radiation. As the orbital period decreases, the black hole inspirals and merges into a more stable end state through the ringdown phase. Before the rise of numerical relativity, perturbative methods were the most effective means of modeling realistic scenarios in GR in the nonlinear regime. 
QNMs describe energy dissipation from fields in a black hole background and can be formally derived from linearized differential equations of GR that constrain perturbations around a black hole solution. QNMs calculated within the linearized framework align closely with those derived from a nonlinear, coupled system of Einstein's equations, particularly for late times \cite{Richartz:2015saa,bishop2016extraction,KokkotasSchmidt:1999,Nollert:1999ji}.

Quasinormal ringing plays a dominant role in phenomena related to black hole perturbations and provides distinct signatures that enable the clear observational identification of hairy GD black holes. To extract substantial information from GW detectors, it is essential to thoroughly understand the main features and signatures of QNMs associated with a hairy GD black hole, whose event horizon behaves as a membrane for classical fields, resulting in a non-Hermitian boundary value problem with complex eigenmodes \cite{BertiCardoso:2009,KonoplyaZhidenko:2011}. The imaginary part of the frequency captures the decay timescale of the black hole perturbation and quantifies the energy lost by the black hole. Perturbed black holes are inherently dissipative due to the effects of the event horizon. These modes dominate the radiation emitted during the intermediate stages of black hole perturbation \cite{Rodrigues:2005yz}. 
 QNMs have a specific dependence on the black hole parameters, making them valuable tools for comparing theoretical predictions with observational data. In this context, the spectrum of hairy GD black holes can unveil possible observational signatures indicative of the GD setup.
Therefore, the main aim of this work is to derive the QN spectrum of frequencies for hairy GD black hole solutions and analyze the resulting deviations from the standard Schwarzschild solution, as the GD hairy parameters are introduced. We will derive the complex frequencies at which black holes oscillate and
at which their GWs propagate. These QNMs contain not only characteristic information about the black hole that emitted them, regarding their no-hair charges, and may also encode primary hair signatures that could be observed in GW detections. We will study the spectra of the QNMs from hairy GD potentials, searching for hair signatures that allow one to distinguish a hairy black hole solution from those satisfying the no-hair conjectures. 

In this work, the GD is implemented for splitting gravitational
sources, generating new terms in the metric that are interpreted as primary hair. We apply perturbation theory to investigate the form and properties of the GWs produced by hairy GD black hole solutions through the analysis of their QNMs. From the Regge--Wheeler and Zerilli equations governing the metric perturbation, which applies a higher-order WKB method, QNMs are obtained and analyzed. We also implement
the same method to obtain the QNMs for a Reissner--Nordstr\"om (RN) black hole with the same values of the outer horizon and square of the charge for each of the three hairy black hole metrics obtained by the GD method, to analyze and quantify their distinctiveness,
which could be considered a primary hair signature in the QNMs of GD hairy black hole GWs.

This paper is organized as follows: Section \ref{secII} is dedicated to reviewing the GD procedure and its main results, leading to GD hairy black hole solutions after some constraints are imposed. In Section \ref{secIV}, we briefly discuss black hole perturbation theory for curved backgrounds and investigate its results when applied to the GD metrics, i.e., the hairy black hole odd potentials. Next, in Section \ref{secV} we present and analyze the quasinormal modes, obtained via sixth-order WKB method, from the GD hairy black hole odd potentials and compare the resulting spectrum with one obtained by the same method when applied to a no-hair solution with the same outer horizon and squared charge values. Section \ref{secV} is devoted to computing the complex frequency differences $\Delta\omega$ between these spectra and analyzing their significance with respect to the error of the WKB method at sixth-order. Conclusions are presented in  Section \ref{secVI}. 

\section{Hairy Black Holes by GD}
\label{secII}
The GD method stems from another procedure known as the minimal geometric deformation (MGD) \cite{Ovalle:2017fgl}, developed to obtain braneworld configurations from general relativistic perfect fluid solutions and conversely \cite{Ovalle:2013vna}. In the context of the MGD, 4-dimensional GR solutions can be derived when terms containing the electric part of the Weyl tensor field and a second-order combination of the energy-momentum tensor are regarded on the right-hand side of the Einstein field equations. These solutions can manifest anisotropy.  In the membrane paradigm of AdS/CFT the additional tensor field, implementing the GD protocol, would contain all the higher-dimensional contributions to the resulting curved geometry through the Weyl tensor field, as Kaluza--Klein modes and moduli fields  \cite{daRocha:2012pt,Abdalla:2009pg,Estrada:2024lhk,Cavalcanti:2016mbe,Contreras:2018vph}. However, even without higher-dimensional scenarios, the 4-dimensional GD apparatus can promote and accommodate eventual effects of dark matter, dark energy, fields beyond baryonic matter, quantum corrections, and  other quantum gravity effects. \clt{In fact, Ref. \cite{Maurya:2024zao} used the GD to investigate Bose--Einstein condensation dark matter models. The possibility of approaching dark matter and dark energy with the GD was also addressed in Ref. \cite{LinaresCedeno:2019aul}. Ref. \cite{Tello-Ortiz:2020ydf} 
   showed that the temporal component of
the extra source $\theta_{\mu\nu}$  mimics the isothermal dark matter
density profile. Refs. \cite{Maurya:2025jzj,Pradhan:2024hne,Yousaf:2024kiv} studied dark matter halos in the context of GD, also describing an extended range of pulsars which have been recently observed. Dark matter signatures in GD compact stellar distributions were proposed in Refs. \cite{Yousaf:2024src,Maurya:2024ylr,Maurya:2025kto}.    
    Besides, since Ref. \cite{Ovalle:2017fgl} showed that the MGD is a particular case of the GD, the MGD was applied in Refs. \cite{ daRocha:2020gee,DaRocha:2019fjr} to study the holographic entanglement entropy of black holes and anisotropic fluids. Also, self-gravitating axion stars were scrutinized through the GD and the MGD. Ref. \cite{Casadio:2023mgl} showed that  MGD axion stars implement mini-massive compact halo objects formed by the condensation of the axion field, representing the final stage of axion miniclusters originated in the cosmic QCD epoch. 
Refs. \cite{darkstars,Ovalle:2013xla,Ovalle:2013vna} proposed physical signatures of holographic braneworlds in MGD black hole models.  
Refs. \cite{Fernandes-Silva:2019fez,Casadio:2016aum}
implemented the black hole quantum portrait, introduced in Ref. \cite{Dvali:2013vxa}, to describe a Bose-Einstein condensate of
 weakly-interacting soft gravitons, 
 encoding quantum effects in Einstein’s classical GR. Ref. \cite{Cavalcanti:2016mbe} explored observational lensing effects of 
MGD black hole solutions in AdS/CFT. Ref. \cite{Meert:2020sqv} probed the GD with the trace and Weyl anomalies, robustly showing that the MGD is consistent with realistic models, in the AdS/CFT
setup. Ref. \cite{daRocha:2017cxu}
used the MGD to investigate glueball dark matter and its collapse into compact stellar configurations, whereas Ref. \cite{Casadio:2017sze} used the MGD to investigate black hole remnants and sub-Planckian black holes in the generalized uncertainty
principle context.}

To implement the GD, the Einstein field equations
\begin{equation}
    G_{\mu\nu} = 8\pi\check{T}_{\mu\nu} 
    \label{EinsteinD}
\end{equation}
are taken into account, 
where a new energy-momentum tensor $\check{T}_{\mu\nu}$ is defined as a combination of two other energy-momentum tensors as
\begin{equation}
    \check{T}_{\mu\nu} \equiv T_{\mu\nu} + \alpha\theta_{\mu\nu}, 
\end{equation}
where $T_{\mu\nu}$ is the usual energy-momentum tensor representing baryonic matter (in this case, a perfect fluid), and $\theta_{\mu\nu}$ is an extra tensor field coupled to $T_{\mu\nu}$. The tensor field $\theta_{\mu\nu}$ represents a second independent gravitational sector, which is related to the primary gravitational sector by the coupling constant $\alpha$ and the following conservation equation \cite{Ovalle:2017fgl,Ovalle:2020kpd}: 
\begin{equation}
 \nabla^{\mu}(T_{\mu\nu} + \alpha\theta_{\mu\nu}) = \nabla^{\mu}\check{T}_{\mu\nu} =0.
 \label{DivFreeFull}
\end{equation}

The seed geometry is obtained by setting $\alpha\to0$, and in the case where it is described by the usual spherically symmetric line element
\begin{equation}
    \mathrm{d}s_{\mathrm{seed}}^2 = -e^{\xi(r)}\mathrm{d}t^2 + e^{\mu(r)}\mathrm{d}r^2 + r^2 \mathrm{d}\Omega^2,
\end{equation}
Eq. \eqref{EinsteinD} then yields
\begin{subequations}
\begin{align}
    8\pi\left( T\indices{^{0}_{0}} + \alpha\theta\indices{^{0}_{0}}\right) &= -\left(r\mu^{\prime} + e^{\mu} - 1\right)\frac{e^{-\mu}}{r^2},\label{00}\\[0.3cm]
        8\pi\left( T\indices{^{1}_{1}} + \alpha\theta\indices{^{1}_{1}}\right) &= -\left(r\xi^{\prime} - e^{\mu} + 1\right)\frac{e^{-\mu}}{r^2}, \label{11}\\[0.3cm]
        8\pi\left( T\indices{^{2}_{2}} + \alpha\theta\indices{^{2}_{2}}\right) &={\left[r\left(\xi^{\prime}\right)^{2} - {\left(r\mu^{\prime} - 2\right)}\xi^{\prime} + 2r \xi^{\prime\prime} - 2\mu^{\prime}\right]} \frac{e^{-\mu}}{4r}.\label{totalEE}
\end{align}
\end{subequations}
The first thing to notice is that by setting the effective energy density, the radial pressure, and the tangential pressures as
\begin{align}
    \check{\rho} &\equiv T\indices{^{0}_{0}} + \alpha\theta\indices{^{0}_{0}},\\
    \check{p}_{r} &\equiv -T\indices{^{1}_{1}} - \alpha\theta\indices{^{1}_{1}},\\
    \check{p}_{t} &\equiv -T\indices{^{2}_{2}} - \alpha\theta\indices{^{2}_{2}},
\end{align}
while assuming $T_{\mu\nu}$ is a perfect fluid, i.e., 
\begin{align}
     T\indices{^{0}_{0}}= \rho,\quad\qquad     T\indices{^{1}_{1}}= -p,\quad\qquad     T\indices{^{2}_{2}}= -p,\label{perfluid}
\end{align}
then Eqs. (\ref{00}) -- (\ref{totalEE}), along with Eq.  \eqref{perfluid}, yields 
\begin{align}
    \Pi &= \check{p}_{r} - \check{p}_{t}\neq  0,
    \label{ani}
\end{align}
where $\rho$ and $p$ are the perfect fluid energy density and isotropic pressure, respectively. Thus, as seen in Eq. \eqref{ani}, a relevant attribute of the GD is the ability to generate GR  solutions for anisotropic fluid configurations \cite{Ovalle:2018ans,Ovalle:2017fgl}. It is important to remember that the existence of the  anisotropy (\ref{ani}) is directly related to the presence of the second gravitational sector implemented by the $\theta_{\mu\nu}$ tensor field.

One could naively try to take Eqs. (\ref{00}) -- (\ref{totalEE}) and separate their sources, constructing two sets of equations, one for $\theta_{\mu\nu}$ and one for $T_{\mu\nu}$. However,  the high non-linearity of Eq. \eqref{EinsteinD} forbids it \cite{Ovalle:2017fgl}. To circumvent this issue, the seed spacetime geometry is modified by the presence of the second gravitational sector. One can show that for a given additional Lagrangian density,  $\mathcal{L}_{\theta}$ associated with $\theta_{\mu\nu}$, the  complete action reads:
\begin{align}
    \hat{S} &= \frac{1}{16\pi}S_{H} + S_{M} + S_{\theta}\\[0.3cm]
            &= \int\left(\frac{1}{16\pi}R + \mathcal{L}_{M} + \mathcal{L}_{\theta}\right)\sqrt{-g}\,\,\mathrm{d}^{4}x,
\end{align}
where $S_{H}, S_{M}$ and $S_{\theta}$ are the Einstein-Hilbert action and the actions regarding the first and second gravitational sectors, respectively. Defining $\theta_{\mu\nu}$ as \cite{Ovalle:2017fgl,Ovalle:2020kpd}
\begin{equation}
    \frac{2}{\sqrt{-g}}\frac{\delta(\sqrt{-g} \mathcal{L}_{\theta})}{\delta g^{\mu\nu}}=2 \frac{\delta\mathcal{L}_{\theta}}{\delta g^{\mu\nu}}-g_{\mu\nu}\mathcal{L}_{\theta}\equiv \theta_{\mu\nu}, 
\end{equation}
  {\color{black}{the resulting geometry is deformed away from the seed solution through the coupling constant $\alpha$, such that:}}
\begin{eqnarray}
    \xi(r) &\mapsto& \nu(r) = \xi(r) + \alpha\color{black} \sigma\left(r\right), \label{decoupling1}\\
    e^{-\mu(r)} &\mapsto& e^{-\lambda(r)} = e^{-\mu(r)} + \alpha\color{black} \kappa\left(r\right),
    \label{decoupling2}
\end{eqnarray}
where $\sigma\left(r\right)$ and $\kappa\left(r\right)$ are the geometric deformation functions \cite{Ovalle:2020kpd}. 
For a spherically symmetric seed solution, the above deformation yields:
\begin{align}
    \mathrm{d}s^2 &= -e^{\nu\left(r\right)}\mathrm{d}t^2 + e^{\lambda\left(r\right)}\mathrm{d}r^2 + r^2 \mathrm{d}\Omega^2, \nonumber\\
    & = -e^{\left[\xi\left(r\right) + \alpha\sigma\left(r\right)\right]}\mathrm{d}t^2 + \cfrac{1}{e^{-\mu\left(r\right)} + \alpha\kappa\left(r\right)}\,\mathrm{d}r^2 + r^2 \mathrm{d}\Omega^2.
       \label{SphDef1}
\end{align}
Note that the $\alpha \rightarrow0$ limits yields $e^{\nu(r)} \mapsto  e^{\xi(r)}$ and $e^{\lambda(r)} \mapsto e^{\mu(r)}$, implying that $\mathrm{d}s^{2} \mapsto\mathrm{d}s_{\mathrm{seed}}^2$, as mentioned previously. {\color{black}{Eq. \eqref{SphDef1} promotes the GD procedure, covered in detail in Refs. \cite{Ovalle:2020kpd, Ovalle:2018umz, Ovalle:2017fgl,Ovalle:2019qyi}. Eqs. \eqref{decoupling1} and \eqref{decoupling2}, when plugged into Eq. \eqref{EinsteinD} yields two sets of equations: one in zeroth order in the coupling constant $\alpha$, which are the Einstein field equations for the seed metric, and a second set of equations at first and at higher orders in $\alpha$, comprising the quasi-Einstein equations. Finding the covariant derivative compatible with Eq. \eqref{SphDef1} and plugging into Eq. \eqref{DivFreeFull} gives us its deformed version, which along with the energy conservation equation to be satisfied by $T_{\mu\nu}$, and an equation of state for the components of $\theta_{\mu\nu}$, leaves us with a definite system that allows us to decouple the sources of gravity. }}

Assuming that the seed metric is a black hole solution, such as the Schwarzschild spacetime, one must impose constraints to make sure that the deformed solution can still describe a black hole, and the consequences will finally lead us to the uncovering of the hairy charges that will later be identified as hair \cite{Ovalle:2020kpd}.

\label{secIII}
From now on, we will work in the context of a Schwarzschild seed spacetime in a tensor-vacuum \cite{Ovalle:2020kpd}, which is achieved by setting $T_{\mu\nu}=0$, leaving a spherically symmetric source $\theta_{\mu\nu}$ as the only energy-momentum tensor present, which will induce the metric deformation. {\color{black}{Solving the equations at zeroth order in $\alpha$ for $\{\xi(r), \mu(r)\}$}}, the line element \eqref{SphDef1} becomes \cite{Ovalle:2020kpd}:
\begin{equation}
    \mathrm{d}s\indices{^{2}} = -\left(1 - \frac{2M}{r}\right)e^{\alpha \sigma\left(r\right)}\mathrm{d}t^{2} + \left({1 - \cfrac{2M}{r} + \alpha \kappa\left(r\right)}\right)^{-1}\mathrm{d}r^{2} + r^{2}\mathrm{d}\Omega^{2},
    \label{SphDef2}
\end{equation}
which is the same as applying the GD method to a seed Schwarzschild spacetime. As seen in Ref. \cite{Ovalle:2020kpd}, to avoid pathological metric signatures, a black hole solution must have the event horizon coincide with Killing horizons, i.e.: 
\begin{equation}
        e^{\nu(r_{\mathrm{h}})} = 0 = e^{-\lambda(r_{\mathrm{h}})},
        \label{causalcond}
    \end{equation}
 where $r=r_{\mathrm{h}}$ is the point that hosts both horizons. Thus, setting:
\begin{equation}\label{eee}
    e^{\nu\left(r\right)} = f\left(r\right) = e^{-\lambda\left(r\right)}, 
\end{equation}
where $f\left(r\right)$ is a generic, static metric function, along with the causal condition in Eq. \eqref{causalcond}, i.e., $f\left(r_{\mathrm{h}}\right)=0$, entails the following relation between the geometric deformation functions \cite{Ovalle:2020kpd}:
\begin{equation}
    \left(1 - \frac{2M}{r}\right)\left(e^{\alpha \sigma\left(r\right)} -1\right) = \alpha \kappa\left(r\right),
\end{equation}
which implies:
\begin{equation}
    f\left(r\right) = \left(1 - \frac{2M}{r}\right)e^{\alpha \sigma\left(r\right)},
    \label{metricfunc}
\end{equation}
consequently, turning 
\begin{equation}
    \mathrm{d}s^2 = -f\left(r\right)\mathrm{d}t^2 + f\left(r\right)^{-1}\mathrm{d}r^2 + r^2\mathrm{d}\Omega^2
    \label{sphlineel2}
\end{equation}
into:
\begin{equation}
    \mathrm{d}s\indices{^{2}} = -\left(1 - \frac{2M}{r}\right)e^{\alpha \sigma\left(r\right)}\mathrm{d}t^{2} + \cfrac{\mathrm{d}r^{2}}{\left(1 - \dfrac{2M}{r}\right)e^{\alpha \sigma\left(r\right)}} + r^{2}\mathrm{d}\Omega^{2}.
    \label{SphDef21}
\end{equation}
As just demonstrated, imposing a well-behaved causal structure yields a direct relation between the deformation functions $\{\sigma(r),\kappa(r)\}$, thus reducing the number of unknown quantities, which now are $\{\sigma\left(r\right), \theta\indices{_{0}^{0}}, \theta\indices{_{1}^{1}}, \theta\indices{_{2}^{2}}\}$. As mentioned before, along with the {quasi-Einstein} equations, an equation of state is needed for $\theta_{\mu\nu}$ to completely solve this system. 
Let us analyze the case where $\theta_{\mu\nu}$ satisfies the dominant energy condition (DEC), implying:
\begin{subequations}
\begin{align}
    \check{\rho} &\geq |\check{p}_{r}|,\label{DECradial}\\
    \check{\rho} &\geq |\check{p}_{t}|. \label{DECtang}
\end{align}
\label{deceq}
\end{subequations}
\!\!\!\clt{Eq. \eqref{eee} therefore infers that 
       $\check{\rho}(r) = -\check{p}_{r}(r)$.}
\clt{Imposing such conditions on the quasi-Einstein equations leads to the following differential equation, as seen in Eq. (84) of Ref. \cite{Ovalle:2020kpd}:
\begin{equation}
    r(r-2M)\left(e^{\alpha\sigma(r)}\right)^{\prime\prime} + 4(r-M)\left(e^{\alpha\sigma(r)}\right)^{\prime} +2e^{\alpha\sigma(r)} - 2 = -\frac{\alpha}{M}\left(r-2M\right)e^{-\frac{r}{M}},
\end{equation}
which can be simplified as:
\begin{equation}
    \left[r(r-2M)\left(e^{\alpha\sigma(r)}-1\right)\right]^{\prime\prime} = \left[\alpha Mre^{-\frac{r}{M}}\right]^{\prime\prime}.
    \label{difeq}
\end{equation}
Integrating twice with respect to $r$  gives
\begin{equation}
    e^{\alpha\sigma(r)} = 1 - \dfrac{1}{\left(r-2M\right)}\left[C_1 +\alpha  Me^{-\frac{r}{M}} + \frac{C_2}{r}\right],
\end{equation}
where $C_{1,2}$ are integration constants, which must be proportional to $\alpha$ so that one can recover the Schwarzschild solution in the limit where $\alpha\rightarrow0$. Setting $C_{1} = \ell_{0}\propto \alpha$ and $C_2 = -Q^2 \propto \alpha$, one obtains \cite{Ovalle:2020kpd}:
\begin{equation}
    e^{\alpha \sigma\left(r\right)} = 1 - \dfrac{1}{(r-2M)}\left[\ell_0 + \alpha Me^{-r/M} - \dfrac{Q^{2}}{r}\right].
    \label{Mfunctions}
\end{equation}
 One will arrive at the explicit form of $Q^{2}$ as a function of $\alpha$ later in this section, after the consequences of the DEC are exhausted. The reason behind this choice of labels for the integration constants becomes evident when one substitutes $e^{\alpha\sigma\left(r\right)}$ back in Eq. \eqref{metricfunc}, turning $f\left(r\right)$ into \cite{Ovalle:2020kpd}}:
\begin{align}
f_{\mathrm{GD}}\left(r\right) &=1-\frac{2\mathcal{M}}{r}+\frac{Q^{2}}{r^{2}}-\alpha\frac{ M}{r} e^{-r/M},
   \label{fGD}
\end{align}
where a re-scaled mass $2\mathcal{M} = 2M + \ell_0$ is introduced. As can be seen, the metric function $f_{\mathrm{GD}}(r)$ (\ref{fGD}) has the same form as $f_{\mathrm{RN}}(r)$, the RN metric function \cite{carroll2019spacetime}, apart from the $e^{-r/M}$ term. It is important to highlight that the dimensionality and the interpretation of the constant $Q$ in $f_{\mathrm{GD}}(r)$ are distinct from the RN counterpart to the hairy GD black hole solutions. In the RN spacetime, $Q$ is the electric charge that fuels the black hole electromagnetic field, whereas in Eq. \eqref{fGD}, $Q$ is proportional to $\alpha$ and has dimensions of length, as does $\ell_0$ \cite{Ovalle:2020kpd}. However, it can be said that both the electromagnetic energy-momentum tensor $E_{\mu\nu}$ and $\theta_{\mu\nu}$ play analogous roles in their respective scenarios, since they both are the agents that deform the seed geometry away from the Schwarzschild solution. As a matter of fact, in Ref. \cite{Ovalle:2018gic}, the tensor $\theta_{\mu\nu}$ is substituted by $E_{\mu\nu}$ to show the effectiveness of the GD method in solving the Einstein field equations and arriving at the RN solution by decoupling the Einstein-Maxwell system.

Due to Eq. \eqref{fGD}, the resulting line element of Eq. \eqref{sphlineel2} yields \cite{Ovalle:2020kpd}:
\begin{align}
    \!\!\!\!\mathrm{d}s^2 \!=\! &-\left(1-\frac{2\mathcal{M}}{r}+\frac{Q^{2}}{r^{2}}-\alpha \frac{M}{r} e^{-r/M}\right)\mathrm{d}t^2+\left(1-\frac{2\mathcal{M}}{r}+\frac{Q^{2}}{r^{2}}-\alpha \frac{M}{r} e^{-r/M}\right)^{-1} \mathrm{d}r^{2} + r^2\mathrm{d}\Omega^2.
    \label{deflineel}
\end{align}
 This solution, for $r\gg M$, goes to an RN solution, as the fourth and extra $1/r$ term in the metric functions falls exponentially with increasing quotient $r/M$. The charges $\{\ell_0, Q\}$ are thus considered to be possible generators of primary hair \cite{Ovalle:2020kpd}, which makes Eq. \eqref{deflineel} a hairy black hole  solution of the Einstein field equations.

\clt{As can be seen in Ref. \cite{Ovalle:2020kpd}, the DEC also implies}:
\begin{equation}
    {Q^2}\geq\frac{r^2\alpha}{4M}\left(r+2M\right)e^{-\frac{r}{M}}.
    \label{Qconstraint1}
    \end{equation}
The line element \eqref{deflineel}, along with the definitions for the effective pressures and energy density, informs that this deformed geometry is generated by:
 \begin{align}\label{41}
        \check{\rho} &= \frac{{Q}^2}{8\pi r^4}-\frac{\alpha e^{-\frac{r}{M}}}{8\pi r^2},\\[0.3cm]
        \check{p}_{r} &= - \frac{{Q}^2}{8\pi r^4}+\frac{\alpha e^{-\frac{r}{M}}}{8\pi r^2} = -\check{\rho},\\[0.3cm]
        \check{p}_{t} &= \frac{{Q}^2}{8\pi r^4}-\frac{\alpha e^{-r/M}}{16\pi Mr}.\label{43}
    \end{align}

Plugging Eqs. (\ref{41}) -- (\ref{43}) into the DEC constraint $\check{\rho}\geq|\check{p}_{t}|$, and simplifying, yields $
    r\geq 2M.$ 
Hence, in this case, satisfying the DEC ultimately means satisfying $r\geq2M$. Furthermore, setting $\ell_0 = \alpha\ell$ and analyzing the line element \eqref{fGD} at the horizon $r=r_{\mathrm{h}}$ shows that:
\begin{equation}
    1-\frac{2M + \alpha\ell}{r_{\mathrm{h}}}+\frac{Q^{2}}{r_{\mathrm{h}}^{2}}-\frac{\alpha M e^{-\frac{r_{\mathrm{h}}}{M}}}{r_{\mathrm{h}}}=0,
    \end{equation}
which means that there will be two horizons: an inner $r_{-}$ and an outer $r_{+}$ horizon, also known as the Cauchy and event horizons, respectively, solutions of the following quadratic equation:
 \begin{equation}
        r_{\mathrm{h}}^2 = -Q^2 +  r_{\mathrm{h}}\left(2M + \alpha\ell + \alpha M e^{-\frac{r_{\mathrm{h}}}{M}}\right).
        \label{eqEH}
    \end{equation}
Eq. \eqref{Qconstraint1} for $r\geq2M$, reads:
\begin{equation}
        Q^2\geq \cfrac{4\alpha M^2}{e^2},
    \end{equation}
which in turn, due to \eqref{eqEH}, yields:
 \begin{equation}
        \ell \geq \frac{M}{e^2},
    \end{equation}
 illustrating the full outspread of the DEC. \clt{It happens that Eq. \eqref{eqEH} is not solvable analytically without additional conditions}. However, three possible analytical solutions are immediate for the following values of $Q^{2}$:
 \begin{subequations}
\begin{align}
    Q^2 &= \cfrac{4\alpha M^2}{e^2},\label{Q1}\\[0.2cm]
    Q^2 &= r_{+}\left(2M + \alpha Me^{-\frac{r_{+}}{M}} \right),\label{Q2}\\[0.2cm]
     Q^2 &= \alpha r_{+}Me^{-\frac{r_{+}}{M}},\label{Q3}
\end{align}     
 \end{subequations}
\clt{which are the very explicit form of $Q$ as a function of the coupling constant $\alpha$,  for each one of the three cases under scrutiny.}
These values, through Eq. \eqref{eqEH}, will yield $r_{+}$ equal to $2M$, $\alpha\ell$ and $2M + \alpha\ell$, respectively. 
To simplify equations, one may set $r_{+} = \beta M$, with $\beta\geq2$, which, along with the Eqs. (\ref{Q1}) -- ( \ref{Q3}), allows to make $M = M\left(Q\right)$, turning $f_{\mathrm{GD}}\left(r\right)$ into:
\begin{equation}
   f_{\mathrm{GD}}\left(r\right) =  1-\frac{2\mathcal{M}}{r}+\frac{Q^{2}}{r^{2}}-\frac{\left({\alpha}/{\beta}\right) r_{+}}{r}\,e^{-\beta\,r/r_{+}}.
    \label{genDef}
\end{equation}
Eq. \eqref{genDef} can be thought of as a master equation, which generates, for each of the three possible pairs $\{r_{+},Q^2\}$, a deformed metric function, presented next in explicit forms:
\begin{align}
    f_{\rm GD_{1}}(r) &=1 - \frac{2\mathcal{M}}{r} + \frac{Q^{2}}{r^{2}} -\frac{Q\sqrt{\alpha} e^{\left(-\frac{2r\sqrt{\alpha} }{eQ} + 1\right)}}{2r}, \label{fgd1} \\[0.5cm]
f_{\rm GD_{2}}(r) &= 1 - \frac{2\mathcal{M}}{r} + \frac{Q^{2}}{r^{2}}  -\frac{Q \alpha e^{\left(-\frac{r}{Q}\sqrt{\frac{{{\left(2e^{\ell_{0}} + \alpha\right)} \ell_{0}}}{{e^{\ell_{0}}}}}\right)}}{\sqrt{\alpha \ell_{0} e^{-\ell_{0}} + 2 \, \ell_{0}}\, r},\label{fgd2}\\[0.5cm]
 f_{\rm GD_{3}}(r) &=  1 - \frac{2\mathcal{M}}{r} + \frac{Q^{2}}{r^{2}} -\frac{Q\alpha e^{\left(-\frac{r}{Q}\sqrt{\frac{{\alpha\left(\ell_{0} + 2\right)}}{{e^{\left(\ell_{0} + 2\right)}}}} + \frac{1}{2}\ell_{0}+ 1\right)}}{\sqrt{\alpha\ell_{0} + 2\alpha}\, r},\label{fgd3}
\end{align}
for $\ell_{0} = \alpha\ell$. \clt{The above equations are the three independent GD metric functions that yield the GD solutions stemming from the three analytical solutions of Eq.\eqref{eqEH}, which will be the core of our analysis from now on.} Applying the following transformation
\begin{align}
    \bm{\bar}{\mathcal{M}} &= \vartheta\mathcal{M},\\[0.3cm]
    \bm{\bar}{Q}^2 &= \vartheta\, Q^2,
\end{align}
where
\begin{equation}
 \vartheta \equiv \left(1 - \cfrac{\alpha e^{-\beta}}{\beta}\right)^{-1},
 \label{transfRN}
\end{equation}
the expression for $r_{+}$ takes the familiar form:
\begin{equation}
r_{+}=\bm{\bar}{\mathcal{M}}+\sqrt{\bm{\bar}{\mathcal{M}}^2-\bm{\bar}{{Q}}^2},
\end{equation}
which reproduces the form that $r_{+}$ takes in the RN geometry \cite{carroll2019spacetime, Ovalle:2020kpd}. It leads to the conclusion that assuming $Q$ to be an electric charge, then it must match a non-linear electrodynamics, with an associated Lagrangian through the P-dual formalism as obtained in Ref. \cite{Ovalle:2020kpd}. However, it is important to remember that $Q^2\propto \alpha$, which has the dimension of length and allows us to recover the Schwarzschild spacetime for $\alpha \rightarrow 0$, as expected.

The quasi-RN nature of the deformed metric functions (\ref{fgd1}) -- (\ref{fgd3}) gives rise to the question of how similar their gravitational wave (GW) signatures are, concerning those from an RN black hole with the same values for $Q^2$. In the next section, we address these questions by computing the QNMs of the deformed metrics and RN black holes with the same $Q^2$ and equivalent mass, so that we can look for possible signatures that would allow us to unequivocally distinguish them  in terms of the gravitational waveform. 

\section{Gravitational Waves from Hairy Black Holes}
\label{secIV}
\subsection{Perturbations in curved backgrounds}
Einstein predicted GWs through a linearized version of GR, an approach that imposes the decomposition of the metric as
\begin{equation}
    g_{\mu\nu} = \mathring{g}_{\mu\nu} + h_{\mu\nu}, ~~~~|h_{\mu\nu}| \ll \mathring{g}_{\mu\nu},\label{fullcurvmetric} 
\end{equation}
where $\mathring{g}_{\mu\nu}$ is the background spacetime and $h_{\mu\nu}$ is a  perturbation maintained small enough by the right choice of gauge provided by GR  invariance under infinitesimal diffeomorphisms  
\cite{Maggiore:2007ulw,BertiCardoso:2009, Ferrari:2020nzo}.

Black hole perturbation theory states that, in the context of curved backgrounds, the perturbed Ricci tensor is given by the Palatini identity \cite{Ferrari:2020nzo,BertiCardoso:2009}: 
\begin{align}
\delta R_{\mu\nu} 
&= \frac{1}{2}\left(\nabla_{\sigma}\nabla_{\mu}h\indices{_{\nu}^{\sigma}} + \nabla_{\sigma}\nabla_{\nu}h\indices{_{\mu}^{\sigma}} - \nabla^{\sigma}\nabla_{\sigma}h_{\mu\nu} - \nabla_{\nu}\nabla_{\mu}h\right),
\end{align}
which yields the perturbed Einstein tensor:
 \begin{equation}
   \delta G_{\mu\nu} \equiv \delta R_{\mu\nu} - \frac{1}{2}\mathring{g}_{\mu\nu}\left(\mathring{g}^{\alpha\beta}\delta R_{\alpha\beta} - h^{\alpha\beta}\mathring{R}_{\alpha\beta} \right) - \frac{1}{2}h_{\mu\nu}\mathring{R}
   \label{pertG}.
\end{equation}
 The previous section presented the GD background with a $1/r^2$ dependency, alike the RN geometry. Hence, to obtain the wavelike equations that govern the GWs emitted from hairy GD solutions,  one may implement the procedure due to Regge--Wheeler \cite{Regge:1957td} and Zerilli \cite{Zerilli:1974ai}. Such a procedure involves performing a harmonic decomposition of the spacetime manifold, decomposing the perturbation tensor, and consequently the perturbed Einstein tensor \cite{Ferrari:2020nzo,BertiCardoso:2009}. Thereafter, one can rewrite them in terms of spherical tensors to take advantage of the spherical symmetry and easily separate the angular from the radial parts by the orthogonality relations. The radial parts may be separated with respect to parity, rendering two sets of Einstein equations: odd and even. Each set, after some manipulation, yields a Schrödinger-like equation known as the Regge--Wheeler and Zerilli master equations for the odd and even perturbations, respectively \cite{Zerilli:1974ai}. In this section, we will focus on the odd perturbations of the hairy GD black hole  solutions. 
 
 The perturbation tensor $h_{\mu\nu}$ may be decomposed with respect to parity as such:
\begin{align}
    h_{\mu\nu}(t,r,\theta,\phi) &= \sum_{m,n}\int^{+\infty}_{-\infty}\tilde{h}^{mn}_{\mu\nu}(\omega,r,\theta,\phi)e^{-i\omega t}\text{d}\omega= \sum_{m,n}\int^{+\infty}_{-\infty} \left( \tilde{h}^{\text{e},mn}_{\mu\nu}+ \tilde{h}^{\text{o},mn}_{\mu\nu}\right)e^{-i\omega t}\text{d}\omega
    \label{paritydecomp}, 
\end{align}
where a Fourier transform was also performed, due to the static character of the hairy GD black hole solutions, and where $\tilde{h}^{\mathrm{e},mn}_{\mu\nu}$ and $\tilde{h}^{\mathrm{o}, mn}_{\mu\nu}$ are the even and odd perturbations, respectively. Their explicit form is given below in terms of scalar and  vector spherical harmonics $\{e^{\mathrm{o},mn}_{a}, Y^{mn}\}$, as well as the blackening factor $f = f\left(r\right)$, the odd $\{\tilde{h}^{\mathrm{o},mn}_{0}, \tilde{h}^{\mathrm{o},mn}_{1}\}$, and even perturbation functions $\{ \tilde{h}^{mn}_{0}, \tilde{h}^{mn}_{1}, \tilde{h}^{mn}_{2}, \tilde{K}^{mn}\}$:
\begin{align}
 \tilde{h}^{\text{o},mn}_{\mu\nu} &=
 \begin{pmatrix}
       0 & 0 & \tilde{h}^{\text{o},mn}_{0}e^{\text{o},mn}_{\theta} & \tilde{h}^{\text{o},mn}_{0}e^{\text{o},mn}_{\phi}\\
       0 & 0 & \tilde{h}^{\text{o},mn}_{1}e^{\text{o},mn}_{\theta} & \tilde{h}^{\text{o},mn}_{1}e^{\text{o},mn}_{\phi}\\
       * & * & 0 & 0\\
       * & * & 0 & 0
    \end{pmatrix},
    \label{oddpertmatrix}\\
    \nonumber\\
    \tilde{h}^{\text{e},mn}_{\mu\nu}&= 
    \begin{pmatrix}
        f\tilde{h}^{mn}_{0} & \tilde{h}^{mn}_{1} & 0 & 0\\
       \tilde{h}^{mn}_{1} & f^{-1}\tilde{h}^{mn}_{2}& 0 & 0 \\
        0 & 0 & r^{2}\tilde{K}^{mn} & 0\\
        0 & 0 & 0 & r^{2}\sin^{2}\theta \tilde{K}^{mn}
    \end{pmatrix}
    Y^{mn}.
    \label{evenpertmatrix}
\end{align} 
The three aforementioned odd equations 
\begin{subequations}
\begin{align}
    \tilde{\beta}^{mn}_{0}\left(\omega, r\right) &= f(\tilde{h}^{\prime\prime}_{0,mn} + i\omega \tilde{h}^{\prime}_{1,mn}) - 2i\omega\frac{f}{r}\tilde{h}^{mn}_{1} + \left[\frac{f^{\prime\prime}}{2} + \frac{n(n+1) + f -1}{r^{2}}\right]\tilde{h}^{mn}_{0},\label{beta0}\\[0.3cm]
    \tilde{\beta}^{mn}_{1}\left(\omega, r\right) &= \frac{i\omega}{rf}\left(r\tilde{h}^{\prime}_{0,mn} - 2\tilde{h}^{mn}_{0}\right) +\left[\frac{f^{\prime\prime}}{2} -\frac{\omega^{2}}{f}+ \frac{\left( n(n+1) - f - 1\right)}{r^{2}}\right] \tilde{h}^{mn}_{1},\label{beta1}\\[0.3cm]
    \check{t}_{mn}\left(\omega, r\right) &= i\omega f^{-1}\tilde{h}^{mn}_{0} + f\tilde{h}^{\prime}_{1,mn} + f^{\prime}\tilde{h}^{mn}_{1}\label{tlm},
\end{align}
\end{subequations}
extracted from the odd Einstein tensor components, when properly manipulated, yield the Regge--Wheeler master equation: 
\clt{\begin{equation}
    \frac{\mathrm{d}^{2} Q_{mn}}{\mathrm{d}r^{2}_{*}} + \left(\omega^{2} - \hat{V}^{\text{odd}}_{mn}\right)Q_{mn} = 0,
    \label{RW1RN}
\end{equation}}
where $Q_{mn}\equiv f\tilde{h}^{mn}_{1}/r$ is the Regge--Wheeler function,  $r_{*}$ is the tortoise coordinate related to $r$ by the equation:
\begin{equation}
    \mathrm{d}r_{*} = \frac{1}{f} \,\mathrm{d}r, 
\end{equation}
and finally, $\hat{V}^{\mathrm{odd}}_{mn}$ is the Regge--Wheeler potential, which governs the behavior of the odd perturbations, constructed according to the version of the Regge--Wheeler equation developed  in Ref. \cite{Zerilli:1974ai} for the RN background:
\begin{equation}
   \hat{V}^{\mathrm{odd}}_{mn}  = \frac{f}{r^2}\left[\frac{1}{2} \, r^2f^{\prime\prime}-rf^{\prime} + {n(n+1)} + f -1 - \frac{2Q^{2}}{r^{2}}\right].
   \label{RNRW}
\end{equation}

By substituting the blackening factor $f$ for the GD metric functions \eqref{fgd1}, \eqref{fgd2}, and \eqref{fgd3}, we obtain the three odd GD potentials $V_{\mathrm{GD_{1}}}$, $V_{\mathrm{GD_{2}}}$ and $V_{\mathrm{GD_{3}}}$, respectively. These potentials will be the source of the QNMs to be computed and analyzed in the next section. Now, let us further analyze the GD metric functions and the corresponding odd potentials they span through Eq. \eqref{RNRW}.
\begin{figure}[ht]
    \centering
    \resizebox{\textwidth}{!}{\input{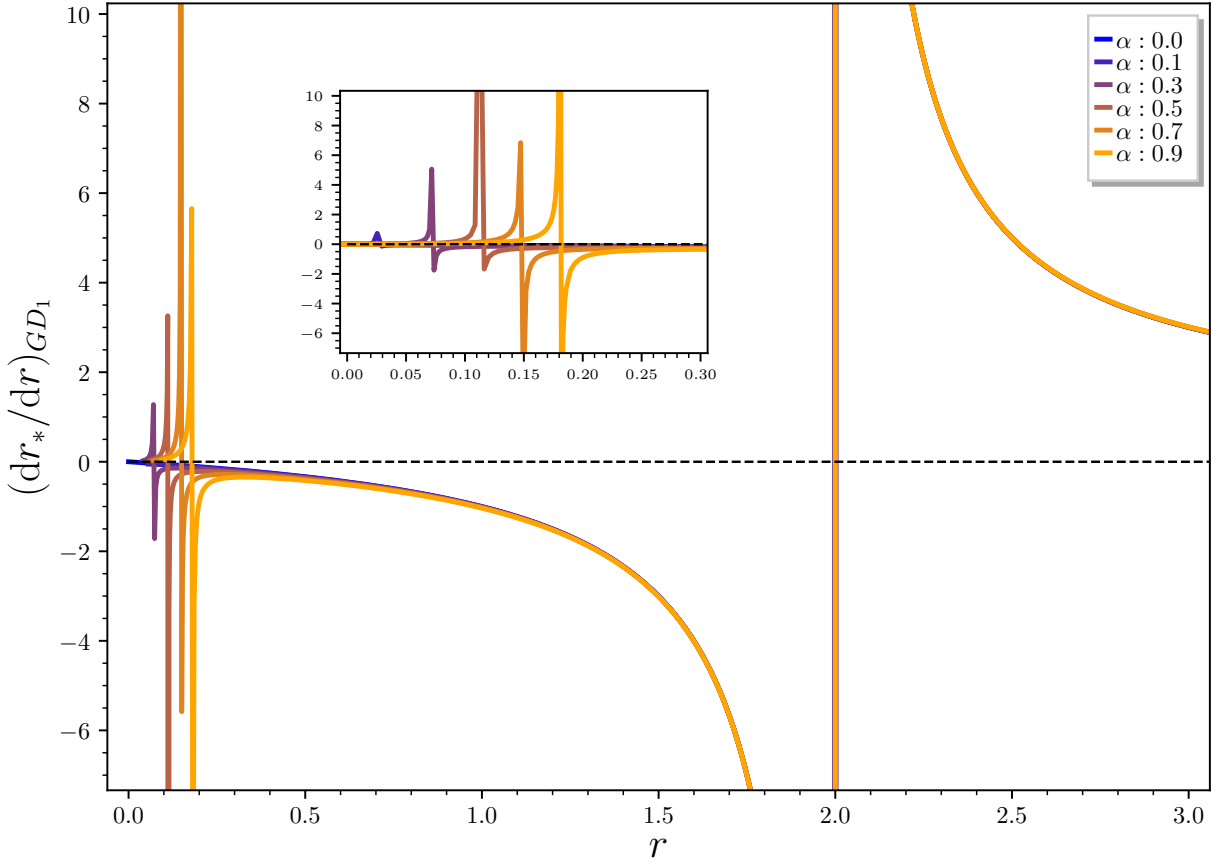}}
    \caption{\footnotesize  Behavior of $\left(\mathrm{d}r_{*}/\mathrm{d}r\right)_{\mathrm{GD_{1}}} = {f^{-1}_{\mathrm{GD_{1}}}}$ as a function of $r$ for different values of $\alpha$, in units of $M$, and $n=2$. The event horizon lies at $r=2$, but the Cauchy horizon assumes larger values for larger $\alpha$.}
    \label{f99}
\end{figure}

\subsection{Hairy black hole odd potentials}
The metric functions of Eqs. \eqref{fgd1}, \eqref{fgd2}, and \eqref{fgd3} may be rewritten in the following, more convenient form:
\begin{subequations}
\begin{align}
    f_{\mathrm{GD_{1}}}\left(r\right) &= 1 - \frac{2 + \alpha\ell + \alpha e^{-{r}}}{r} + \frac{4\alpha e^{-2}}{r^2}\,\,\,\,\,\left(r_{+} = 2\right)\,\,\,,\label{fgd1m1}\\[0.3cm]
    f_{\mathrm{GD_{2}}}\left(r\right) &= 1 - \frac{2 + \alpha\ell + \alpha e^{-{r}}}{r} + \frac{\alpha\ell\left(2 + \alpha e^{-\alpha\ell}\right)}{r^2}\,\,\,\,\,\left(r_{+} = \alpha\ell\right)\,\,\,,\label{fgd2m1}\\[0.3cm]
    f_{\mathrm{GD_{3}}}\left(r\right) &=  1 - \frac{2 + \alpha\ell + \alpha e^{-{r}}}{r} + \frac{\alpha\left(2 + \alpha\ell\right) e^{-\alpha\ell - 2}}{r^2}\,\,\,\,\,\left(r_{+} = 2+\alpha\ell\right)\,\,\,,\label{fg3m1}
\end{align}
\end{subequations}
where we substituted the corresponding values for $Q^2$ from Eqs. \eqref{Q1}, \eqref{Q2}, and \eqref{Q3}, respectively,  and set $M = 1$. For the rest of the discussion, we set $r_{+}=3$ for $f_{\mathrm{GD_{2}}}$ and $f_{\mathrm{GD_{3}}}$, such that they satisfy the DEC.

\begin{figure}
    \centering
    \resizebox{\textwidth}{!}{\input{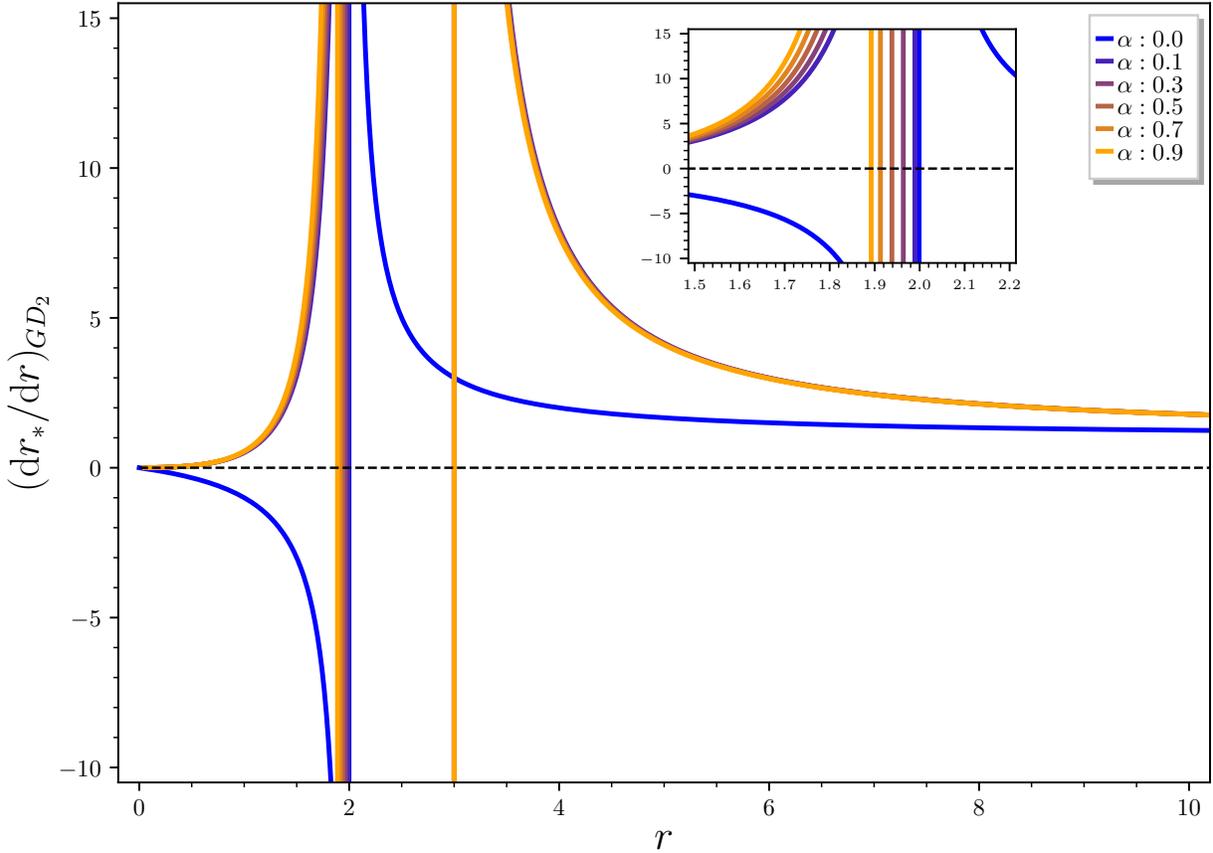}}
    \caption{\footnotesize  Behaviour of $\left(\mathrm{d}r_{*}/\mathrm{d}r\right)_{\mathrm{GD_{2}}} = {f^{-1}_{\mathrm{GD_{2}}}}$ as a function of $r$ for different values of $\alpha$, in units of $M$, and $n=2$. The event horizon is always at $r=3$, except for $\alpha=0$, but the Cauchy horizon assumes smaller values for larger $\alpha$, where $\alpha=0$ corresponds to the Schwarzschild spacetime.}
    \label{f101}
\end{figure}

\begin{figure}
    \centering
    \resizebox{\textwidth}{!}{{\input{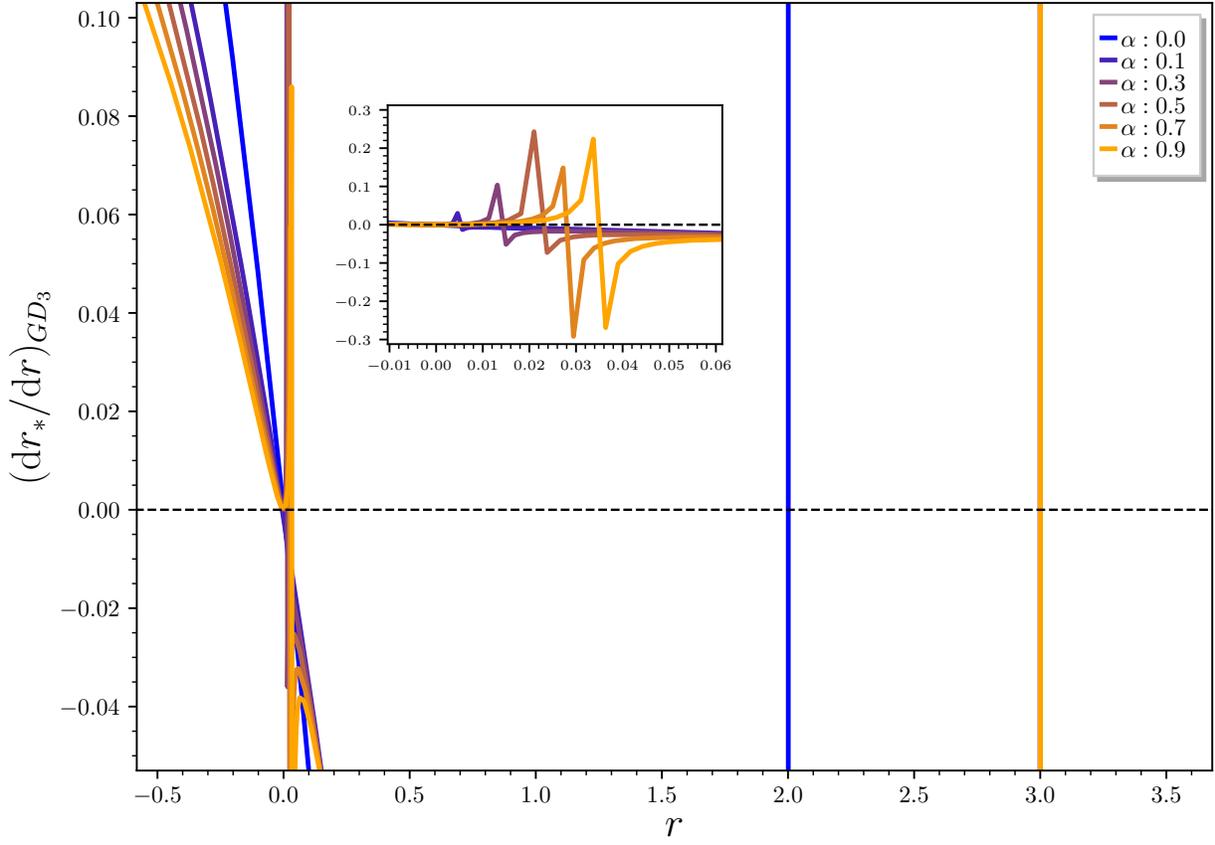}}}
    \caption{\footnotesize  Behaviour of $\left(\mathrm{d}r_{*}/\mathrm{d}r\right)_{\mathrm{GD_{3}}} = {f^{-1}_{\mathrm{GD_{3}}}}$ as a function of $r$ for different values of $\alpha$, in units of $M$, and $n=2$. The event horizon is always at $r=3$, except for $\alpha=0$, but the Cauchy horizon assumes larger values for larger $\alpha$, where $\alpha=0$ corresponds to the Schwarzschild spacetime.}
    \label{f103}
\end{figure}

In Figs. \ref{f99} --  \ref{f103}, we see that $r_{\mathrm{h}}$ indeed assumes two values in all three potentials. In these plots, the outer horizons $r_{+}$ are fixed either at $r=2$ or $r=3$, in units of $M$, for all values of $\alpha$, while the inner horizons $r_{-}$ assume different values for different $\alpha$. In Fig. \ref{f99} we see that, for $\alpha>0$, the metric functions follow the behavior of the Schwarzschild solution, here represented by the $\alpha=0$ line. However, as we approach $r=0$ from the right, the metric functions for $\alpha>0$ deviate from $\alpha=0$, and it becomes clear that the values for $r_{-}$ grow with increasing $\alpha$, approaching $r_{+}=2$.

As seen in Fig. \ref{f101}, for every $\alpha>0$, the metric function  $f_{\mathrm{GD_{2}}}$ reproduces the behavior of the usual Schwarzschild blackening factor $f_{\mathrm{Schw}}$, asymptotically, but deviates from it as one approaches $r=0$ from $+ \infty$, alike the previous case. However, in contrast to $f_{\mathrm{GD_{1}}}$, in this case the values for $r_{-}$ decrease with increasing $\alpha$, making the inner horizons increasingly distant from $r_{+}$, here fixed at $r=3$. As for $f_{\mathrm{GD_{3}}}$, its behavior is very similar to $f_{\mathrm{GD_{2}}}$; the key difference is, again, the response of the inner horizons. The event horizon is
again fixed at $r = 3$, but the values of $r_{-}$ grow with increasing $\alpha$, approaching outer and inner horizons, as it did for $f_{\mathrm{GD_{1}}}$.

\begin{figure}
    \centering
\resizebox{\textwidth}{!}{\input{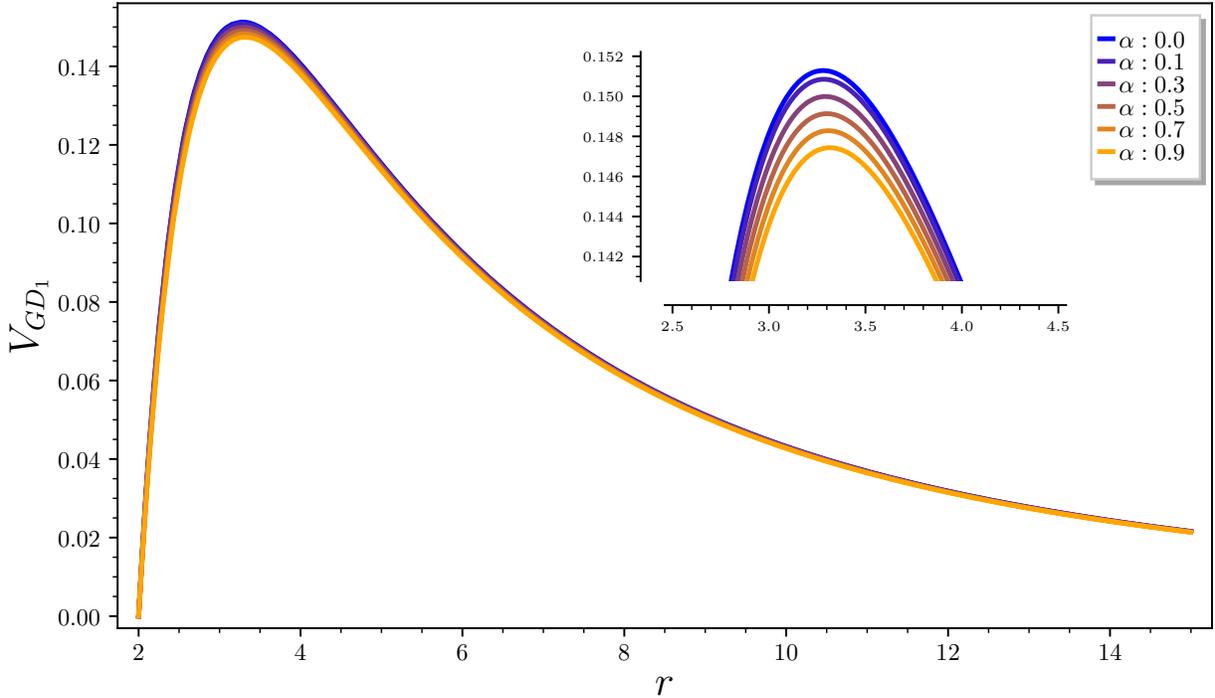}}
    \caption{\footnotesize  Potential $V_{\mathrm{GD_{1}}}$ as a function of $r$, in units of $M$, for multiple values of $\alpha$, with fixed $r_{+} = 2$.}
    \label{V99g}
\end{figure}

\begin{figure}
    \centering
    \resizebox{\textwidth}{!}{\input{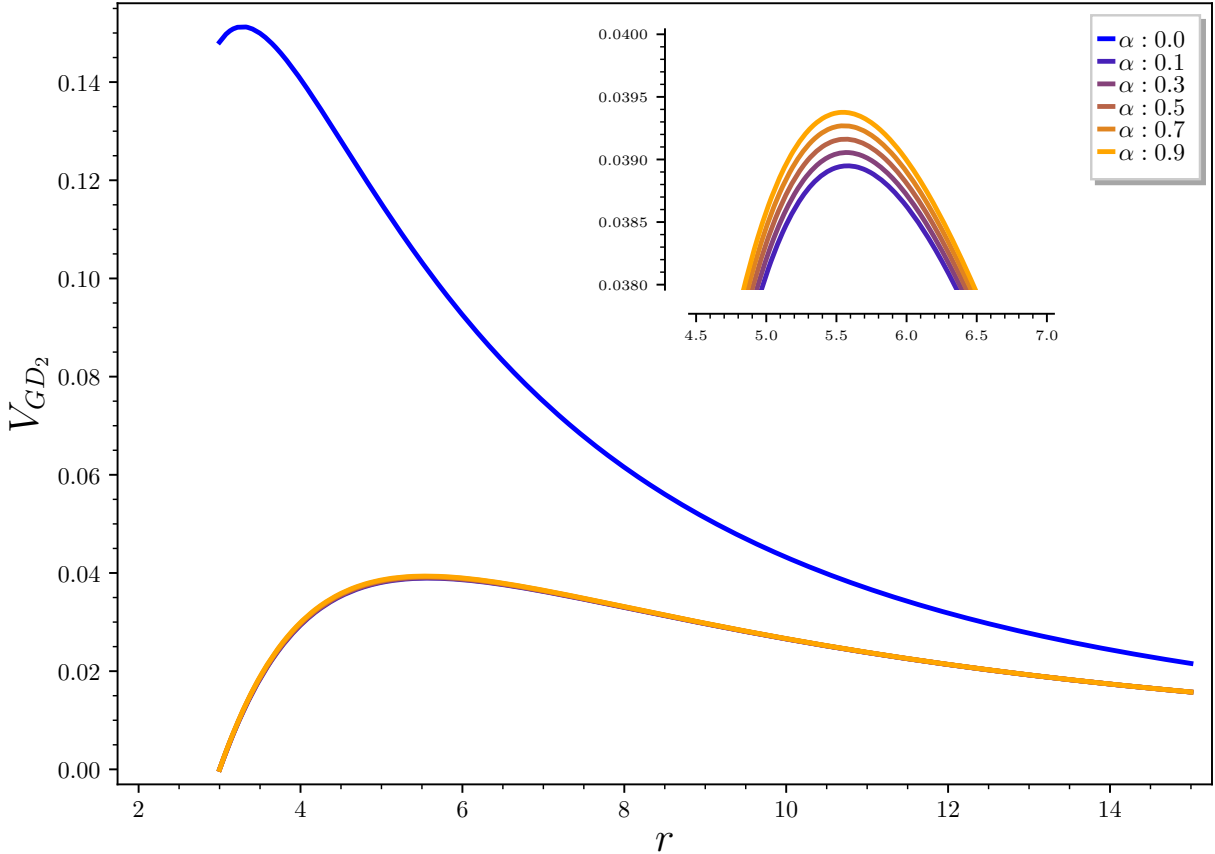}}
    \caption{\footnotesize  Potential $V_{\mathrm{GD_{2}}}$ as a function of $r$, in units of $M$, for multiple values of $\alpha$, with fixed $r_{+} = \alpha\ell =  3$.}
    \label{V101g}
\end{figure}

\begin{figure}
    \centering
    \resizebox{\textwidth}{!}{\input{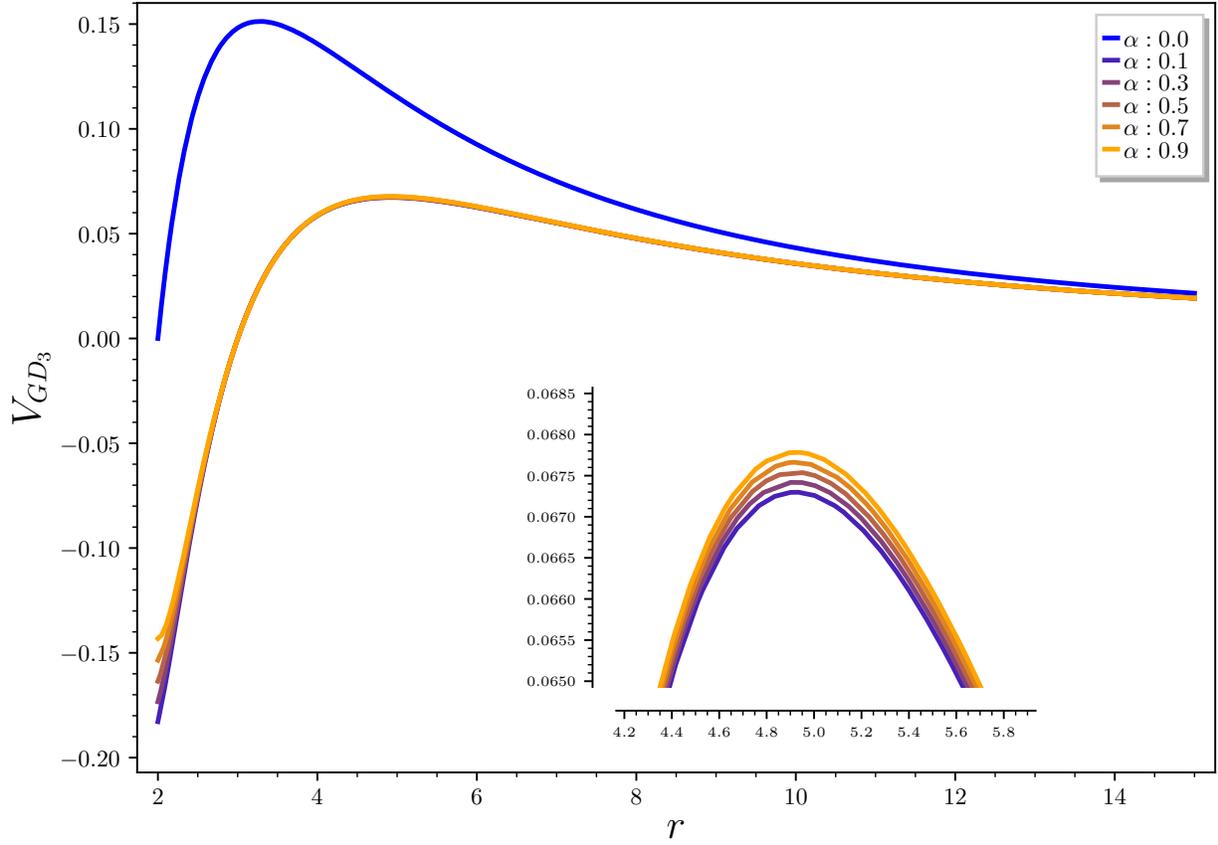}}
    \caption{\footnotesize  Potential $V_{\mathrm{GD_{3}}}$ as a function of $r$, in units of $M$, for multiple values of $\alpha$, with fixed $r_{+} = 2 + \alpha\ell =  3$.}
    \label{V103g}
\end{figure}

As previously mentioned, the $1/r^2$ terms indicate that we must use the Regge--Wheeler equation developed for RN spacetimes to appropriately understand the odd perturbations for the hairy black hole  solutions. Thus,  plugging the deformed metric functions into Eq. \eqref{RNRW} we obtain the full GD potentials, as seen in Figs. \ref{V99g} --  \ref{V103g}, where it is  also possible to see that for $\alpha\rightarrow0$, all three potentials $\big\{V_{\mathrm{GD_{1}}}, V_{\mathrm{GD_{2}}}, V_{\mathrm{GD_{3}}}\big\}$ become
\begin{equation}
    V = {\left({1} - \frac{2}{r}\right)}\left(\frac{n(n+1)}{r^2}  - \frac{6}{r^{3}} \right),
\end{equation}
which is exactly the Regge--Wheeler potential for a  Schwarzschild background, as expected, since in this limit $f_{\mathrm{GD}}\left(r\right)$ reverts to the Schwarzschild metric function $f_{\mathrm{Schw}}\left(r\right)$. Looking at Fig. \ref{V99g}, it is possible to see this smooth transition as $\alpha$ bounces from the infimum to the supremum of the interval $[0.0,0.9]$. The behavior of the GD and the seed potentials is essentially the same, but  higher values of $\alpha$ yield lower maximum values in the GD potentials. In other words, the presence of $\alpha$ diminishes the potential barrier, which assumes its higher maximum value for $\alpha=0$, i.e., in the Schwarzschild case. 

The transition to the seed solution is not always monotonic. 
Take the potentials $V_{\mathrm{GD_{2}}}$ and $V_{\mathrm{GD_{3}}}$  in Figs. \ref{V101g} and \ref{V103g}, respectively. The behavior for $\alpha>0$ starts the same but deviates more and more as it approaches $r\to3$ from $+\infty$. Starting at $\alpha=0$, the next increment in $\alpha$ leads to the lowest maximum values of the GD potentials. Subsequent increments elevate the potential barriers, with a peak at $\alpha=0.9$, in the $\alpha>0$ regime. Thus, starting at $\alpha=0.9$, the maximum values of $V_{\mathrm{GD_{2}}}$ and $V_{\mathrm{GD_{3}}}$ decrease as $\alpha$ decreases, but as soon as $\alpha$ reaches zero, they jump to the Schwarzschild case, where they assume their highest possible value. Hence, the maximum values of $V_{\mathrm{GD_{2}}}$ and $V_{\mathrm{GD_{3}}}$, for all $\alpha>0$ will be lower with respect to the Schwarzschild potential. However, between GD potentials, higher $\alpha$ values yield higher potential maximum values. Hence, even with different responses for $\alpha\in [0.1, 0.9]$, all three GD potentials agree with the fact that $\alpha$ effectively lowers the potential barrier.

\section{QNMs of GD Hairy Black Holes}
\label{secV}
In this section, the QNMs modes from the GD potentials will be computed and analyzed. 
\subsection{Hairy black hole QNM  frequencies} Hereon we implement the sixth-order WKB approximation \cite{Avalos:2023jeh, Konoplya:2019hlu, Konoplya:2019hlu} to derive the QNMs from the GD potentials. The complex frequencies in this method are given by 
\begin{equation}
    \omega^2 = V_{0} - \frac{i}{2}{\sqrt{V^{\prime\prime}_{0}}}\left( \Lambda_{2} + \Lambda_{3} + \Lambda_{4} + \Lambda_{5} + \Lambda_{6} + n_{0} + \frac{1}{2} \right), 
    \label{omega2}
\end{equation}
where $V_{0}$ represents the potential evaluated at its maximum, $V^{\prime\prime}_{0}$ represents its second derivative also evaluated at its maximum, and where $\Lambda_{i}$ are the correction factors, which explicit forms can be found in Ref. \cite{KonoplyaZhidenko:2011}.

\begin{figure}[h!]
 \centering
\includegraphics[width=0.95\textwidth]{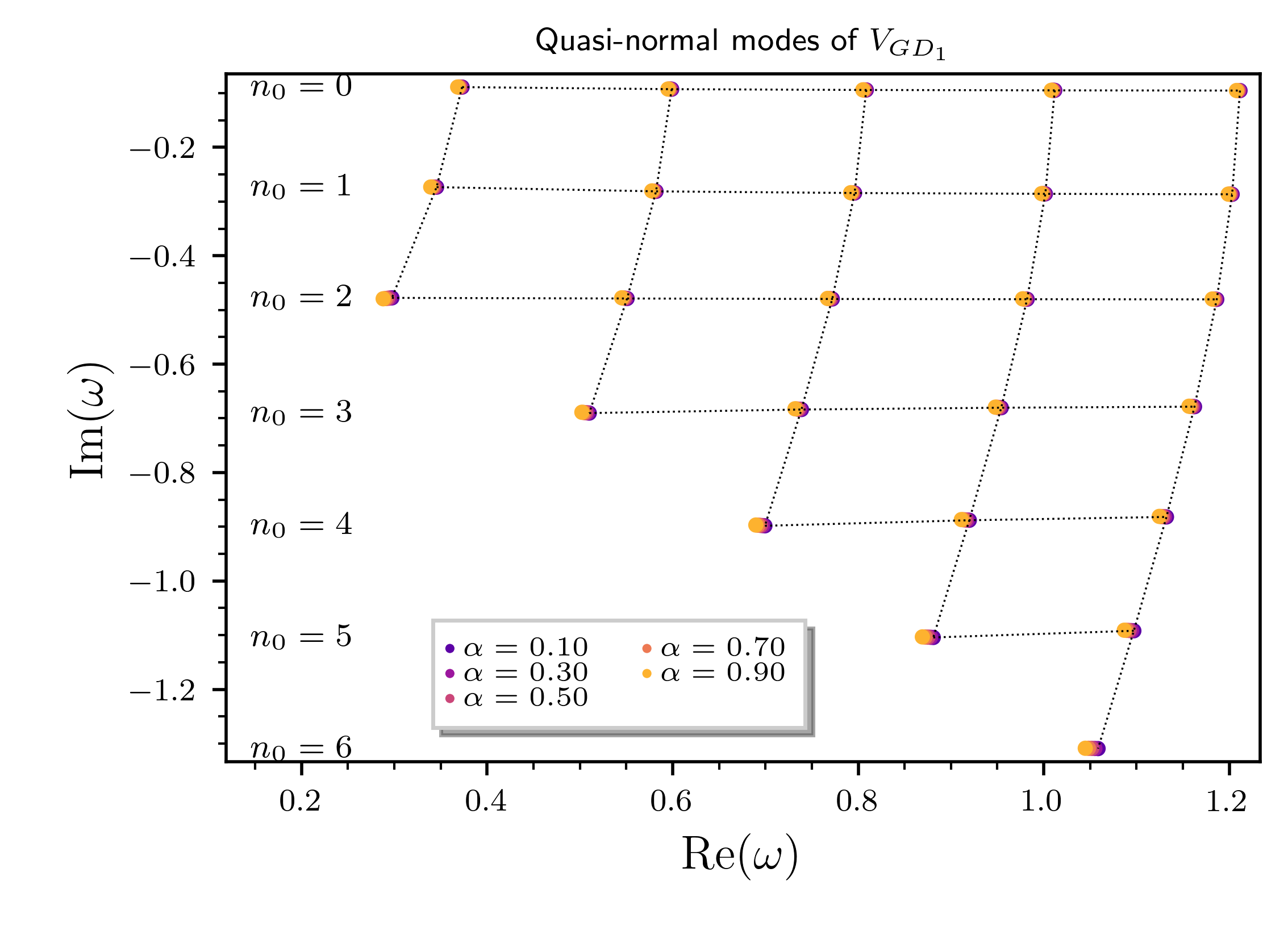}
 \caption{\footnotesize QNMs of the potential $V_{\mathrm{GD_{1}}}$ for a range of values of $\alpha$. The vertical dotted lines correspond, from left to right, to the harmonic numbers $n=2,3,4,5$ and $6$ respectively. The horizontal lines correspond to the displayed overtone.}\label{plotV_1}
\end{figure}

Taking the necessary derivatives of $V_{\mathrm{GD_{1}}}$ and  plugging them into the square root of Eq.  \eqref{omega2} gives us the QNMs depicted in Fig. \ref{plotV_1} and displayed in Table \ref{qnmVGD1} for multiple values of harmonics $n$, overtone number $n_{0}$ and \clt{for $\alpha\in \{0.1, 0.3, 0.5, 0.7, 0.9\}$}, maintaining $n_{0}\leq n$, where the method is suggested to be more accurate \cite{KonoplyaZhidenko:2011,Konoplya:2018ala, Churilova:2019qph, Rincon:2019jal, Rincon:2020iwy}. In this case, one has a fixed $\ell = e^{-2}$, and the first thing to notice is that for fixed values of  $\alpha$ and harmonic $n$, the imaginary part of the modes increases with the overtone number $n_{0}$, as expected from the definition of the overtones \cite{Schutz:1985km,KonoplyaZhidenko:2011}, while the real part decreases. For fixed $\{\alpha, n_{0}\}$, the absolute values of both real and imaginary parts increase for higher harmonics. However, for fixed $\{n, n_{0}\}$, $|\mathrm{Im}\left(\omega\right)|$ decreases with higher values of $\alpha$, as does its real part; ergo, the coupling constant $\alpha$ slows the decay of the overtones. Therefore, the solutions given by $f_{\mathrm{GD_{1}}}$ for each $\alpha>0$ oscillate at a slower rate, but will oscillate longer than a Schwarzschild black hole. This result is similar to the ones obtained for scalar perturbations in Refs. \cite{Cavalcanti:2022cga, Avalos:2023jeh} with sixth and higher orders of the WKB method for the same hairy metric function $f_{\mathrm{GD}_{1}}$.

\begin{figure}[h!]
 \centering
\includegraphics[width=0.95\textwidth]{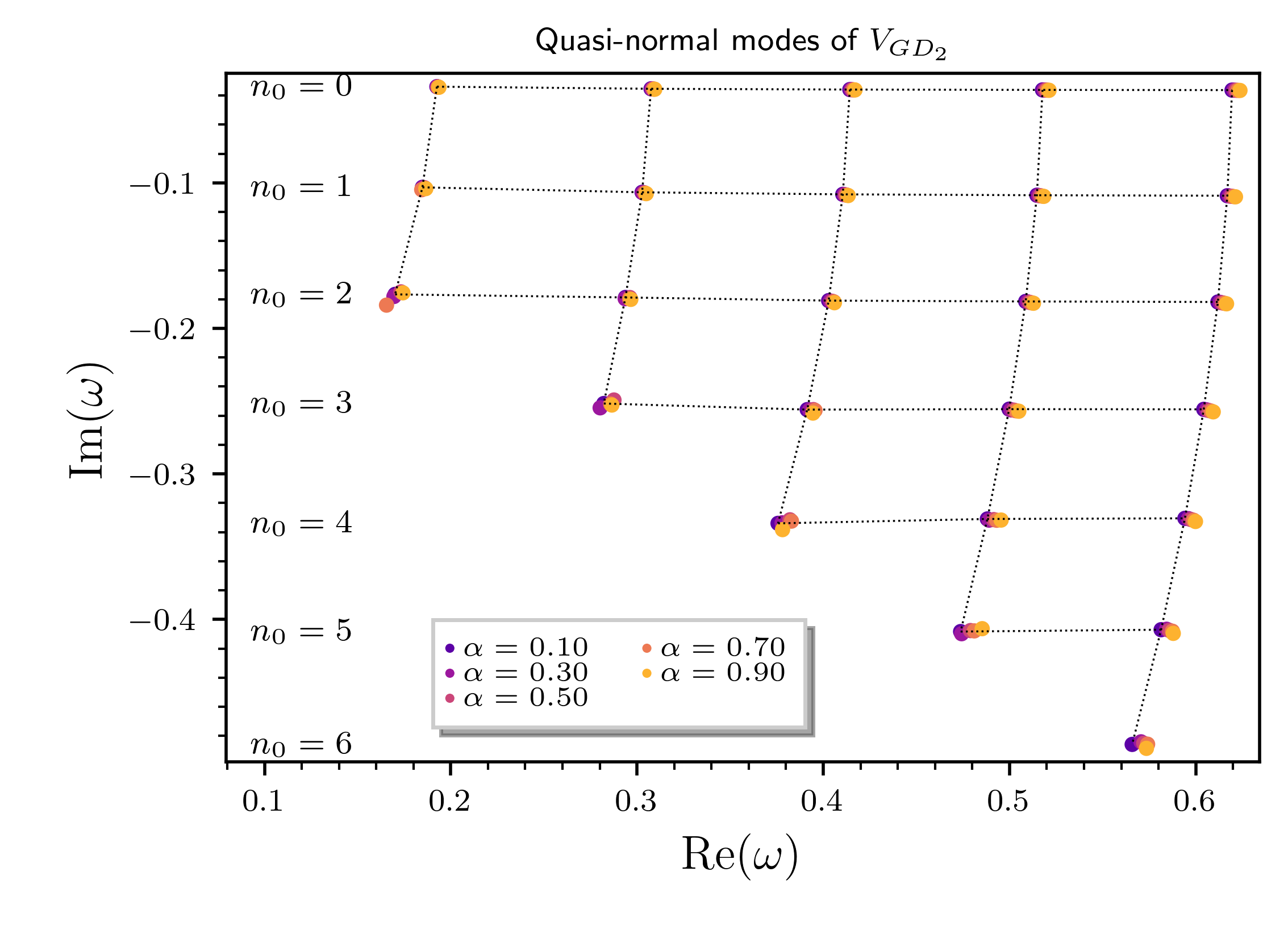}
 \caption{\footnotesize QNMs of the potential $V_{\mathrm{GD_{2}}}$ for a range of values of $\alpha$. The vertical dotted lines correspond, from left to right, to the harmonic numbers $n=2,3,4,5$ and $6$ respectively. The horizontal lines correspond to the displayed overtone.}\label{plotV_2}
\end{figure}

The complex frequencies obtained for $f_{\mathrm{GD_{2}}}$ are displayed in Fig. \ref{plotV_2} and Table \ref{qnmVGD2} for the same range of values of $\{\alpha, n, n_{0}\}$ used for $f_{\mathrm{GD}_{1}}$, and fixed event horizon $r_{+}=\alpha\ell = 3$, in units of $M$. In this set, we encounter the same results for fixed $\alpha$, but in contrast with $V_{\mathrm{GD}_{1}}$, the damping grows with the increase of $\alpha$. Therefore, GD solutions by $f_{\mathrm{GD}_{2}}$ for $\alpha>0$ seem to oscillate and fade more rapidly with higher values of the $\alpha$ parameter. 

\begin{figure}[h!]
 \centering
\includegraphics[width=0.95\textwidth]{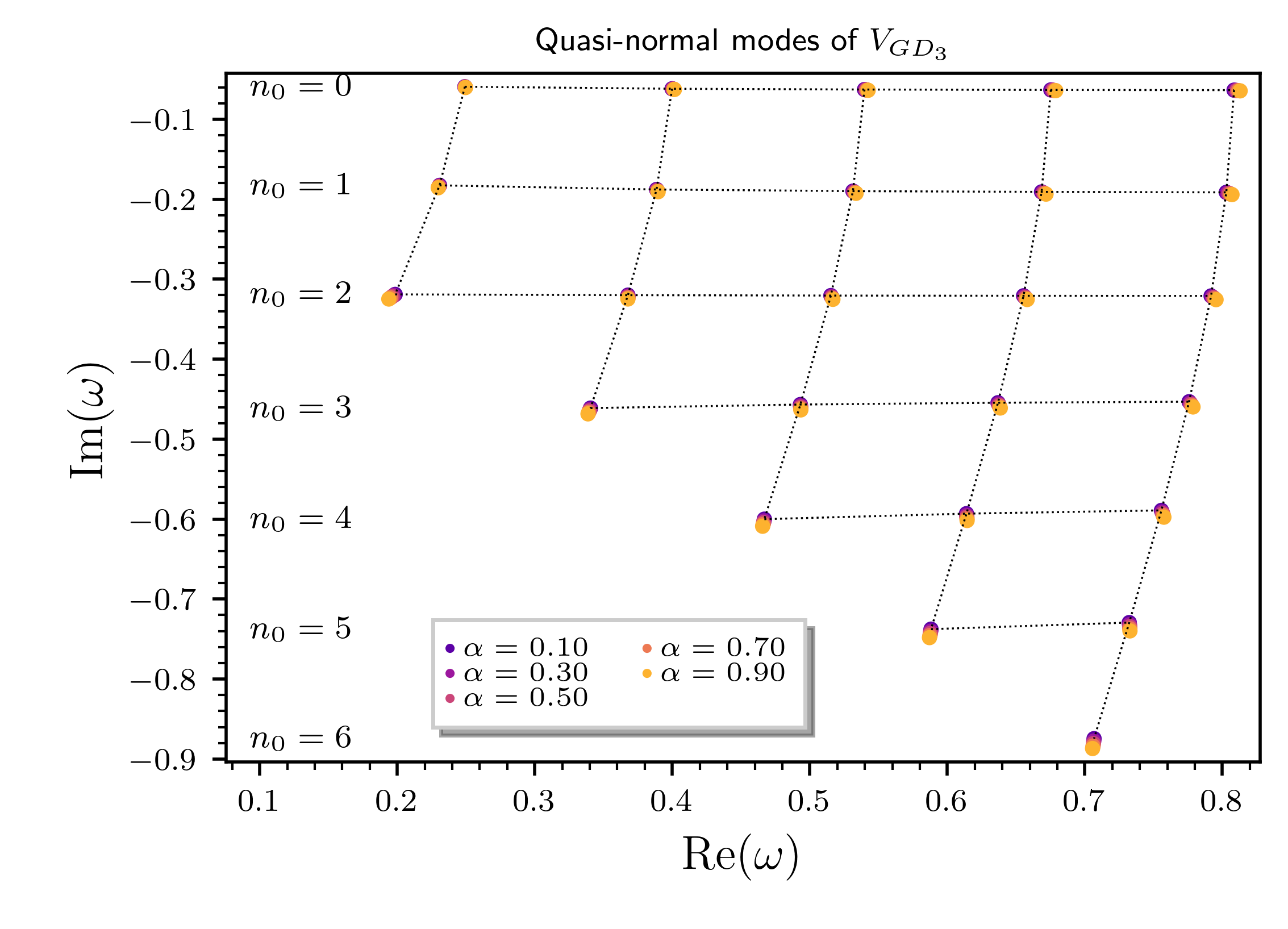}
 \caption{\footnotesize QNMs of the potential $V_{\mathrm{GD_{3}}}$ for a range of values of $\alpha$. The vertical dotted lines correspond, from left to right, to the harmonic numbers $n=2,3,4,5$ and $6$ respectively. The horizontal lines correspond to the displayed overtone.}\label{plotV_3}
\end{figure}

As for $f_{\mathrm{GD_{3}}}$, the behaviour of its QNMs, seen in Fig. \ref{plotV_3} and Table \ref{qnmVGD3}, is very similar to those of $f_{\mathrm{GD_{2}}}$, with some distinctive traits. For fixed $\{\alpha, n\}$, the real part decreases, while the imaginary part increases for higher $n_{0}$, once again, as expected. For fixed $\{n, n_{0}\}$, as obtained for $f_{\mathrm{GD}_{2}}$, the imaginary part increases in absolute value as $\alpha$ increases. The real part of the frequencies increases in the overall range \clt{$\alpha\in \{0.1, 0.3, 0.5, 0.7, 0.9\}$} for lower overtones, but decreases for higher overtones, which is the region where the method has lower accuracy. Thus, this anomaly may be related to shortcomings of the method or even due to the order implemented. 

For both $V_{\mathrm{GD_{2}}}$ and $V_{\mathrm{GD_{3}}}$, a drop in range for both parts of the complex frequencies with respect to the Schwarzschild case is observed. This gap is slightly smaller for $V_{\mathrm{GD_{3}}}$, compared with the one observed for $V_{\mathrm{GD_{2}}}$, which may be caused by the different forms of the GD potentials, or due to the different values of $\alpha\ell$ used. Since we chose to perform our analysis in a regime of a fixed event horizon, in order to keep $r_{+}=3$, $\alpha\ell = \ell_0$ must equal $3$ for $f_{\mathrm{GD_{2}}}$ and equal $1$ for $f_{\mathrm{GD_{3}}}$. Given what has been observed so far from $\alpha$, it is reasonable to expect the gap between the deformed potentials and the Schwarzschild case to increase as $\alpha$ increases, for fixed $\ell$. 

Our initial analysis of the GD potentials showed a clear distinction between their maximum values, in comparison with the Schwarzschild odd potential. Despite the transition from the hairy to the no-hair values being monotonic for $V_{\mathrm{GD_{1}}}$, but not for $\{V_{\mathrm{GD_{2}}}, V_{\mathrm{GD_{3}}}\}$, all three GD potentials presented lower maximum values, for all values of $\alpha$ different than zero. This feature is carried through the calculations of the WKB method and reflected in the complex frequencies, especially in their imaginary parts. Therefore, our analysis has led us to conclude that, despite their particularities inside the $\alpha>0$ regime, GD hairy black holes satisfying the DEC have lower damping rates than a Schwarzschild black hole.

\subsection{The spectrum of no-hair black holes with the same $\{r_{+}\,,\, Q^2\}$}

A natural question to ponder is whether the spectrum just obtained is exclusive to the GD hairy black hole solutions presented. In other words, is it possible for a no-hair solution to reproduce the same spectrum with some combination of its parameters? To start addressing this issue, we call back to Eq. \eqref{transfRN}, from which we conclude that both RN and GD spacetimes cannot have the same values for all three parameters $\{r_{+}\,,\, M\,,\, Q^2\}$ while having $\alpha\neq0$. So, \clt{to compare GD and RN solutions}, one must choose a combination of two out of the three parameters. Here we will work with RN black holes with equivalent values for the square of its charge $Q^2$, and outer horizon $r_{+}\in \{2,3\}$, from the GD potentials. \clt{Since there are three values of $Q^{2}$ for fixed $r_{+}$, given by Eqs. \eqref{Q1} -- \eqref{Q3}, we will compute three other QNM spectra, only now for the corresponding RN with $r^{\mathrm{RN}}_{+} = r_{+}$ and $Q^2_{\mathrm{RN}} = Q^{2}$ for each of the aforementioned values of $\{r_{+}, Q^{2}\}$ . With said spectra for the RN and GD black holes with the same $\{r_{+}\,,\,Q^2\}$, we will be able to draw an objective comparison and analyze how different they are.} 

For the metric function $f_{\mathrm{RN}}$, the outer horizon is given by:
\begin{equation}
    r_{+} = \bm{\bar}{M} + \sqrt{\bm{\bar}{M}^2 - \bm{\bar}{Q}^2}, 
\end{equation}
where $\{\bm{\bar}{M}, {\bm{\bar}{Q}}\}$ are the mass and electric charge of the no-hair RN spacetime. Solving for $\bm{\bar}{M}$, we get:
\begin{equation}
   2\bm{\bar}{M} = \frac{\bm{\bar}{Q}^2}{r_{+}} + r_{+}.
\end{equation}

For $\bm{\bar}{Q}^2 = Q^2$, from Eqs. \eqref{Q1}, \eqref{Q2}, and \eqref{Q3}, we obtain:
\begin{align}
    \bm{\bar}{M}_{1} &= \alpha e^{-2} + 1, \label{Mrn1}\\[0.3cm]
    \bm{\bar}{M}_{2} & = \frac{1}{2}\left(5+\alpha e^{-3}\right), \label{Mrn2}\\[0.3cm]
    \bm{\bar}{M}_{3} &= \frac{1}{2}\left(3 + \alpha e^{-3}\right), \label{Mrn3}
\end{align}
where we substituted $r_{+}=2$ for $\bm{\bar}{M}_{1}$, and $r_{+}= 3$ for $\bm{\bar}{M}_{2}$ and $\bm{\bar}{M}_{3}$.

Substituting Eqs. (\ref{Mrn1}) --  \eqref{Mrn3} back into $f_{\mathrm{GD_{1}}}$, $f_{\mathrm{GD_{2}}}$, and $f_{\mathrm{GD_{3}}}$, respectively, and plugging the new metric functions into Eq. \eqref{RNRW}, gives us the odd potentials $V_{\mathrm{RN_{1}}}$, $V_{\mathrm{RN_{2}}}$ and $V_{\mathrm{RN_{3}}}$. These are odd-RN potentials, with same values for $r_{+}$ and charge $Q^2$ as $V_{\mathrm{GD_{1}}}$, $V_{\mathrm{GD_{2}}}$ and $V_{\mathrm{GD_{3}}}$, respectively. Their QNMs are obtained by taking the necessary derivatives and substituting them into Eq. \eqref{omega2}, which will then yield their complex frequencies. Our goal is to measure how close these frequencies are to the ones we obtained for the GD potentials. To this end, we define:
\begin{equation}
    \Delta\omega \equiv |\omega_{\mathrm{\scriptsize GD}}- \omega_{\mathrm{\scriptsize RN}}|,
    \label{DeltaOmega}
\end{equation}
which we obtain for each pair $\{V_{\mathrm{GD_{i}}}\,,\,V_{\mathrm{RN}_{i}}\}$, for $i = \{1,2,3\}$, i.e., a GD potential and a correspondent RN potential with the same values for $r_{+}$ and $Q^2$. The results depicted in Figs. \ref{plotDV_1} --  \ref{plotDV_3} and which compose the tables of Appendix \ref{AppendixOmega} tell us that, for all three pairs $\{V_{\mathrm{GD_{i}}}\,,\,V_{\mathrm{RN_{i}}}\}$, the absolute value $|\Delta$Im$\left(\omega\right)|$  increases as $\alpha$ increases, i.e., the imaginary parts are more distinguished from each other for higher values of the primary hair $\alpha$, for a given pair $\{n, n_{0}\}$.

\begin{figure}[ht]
 \centering
\includegraphics[width=0.9\textwidth]{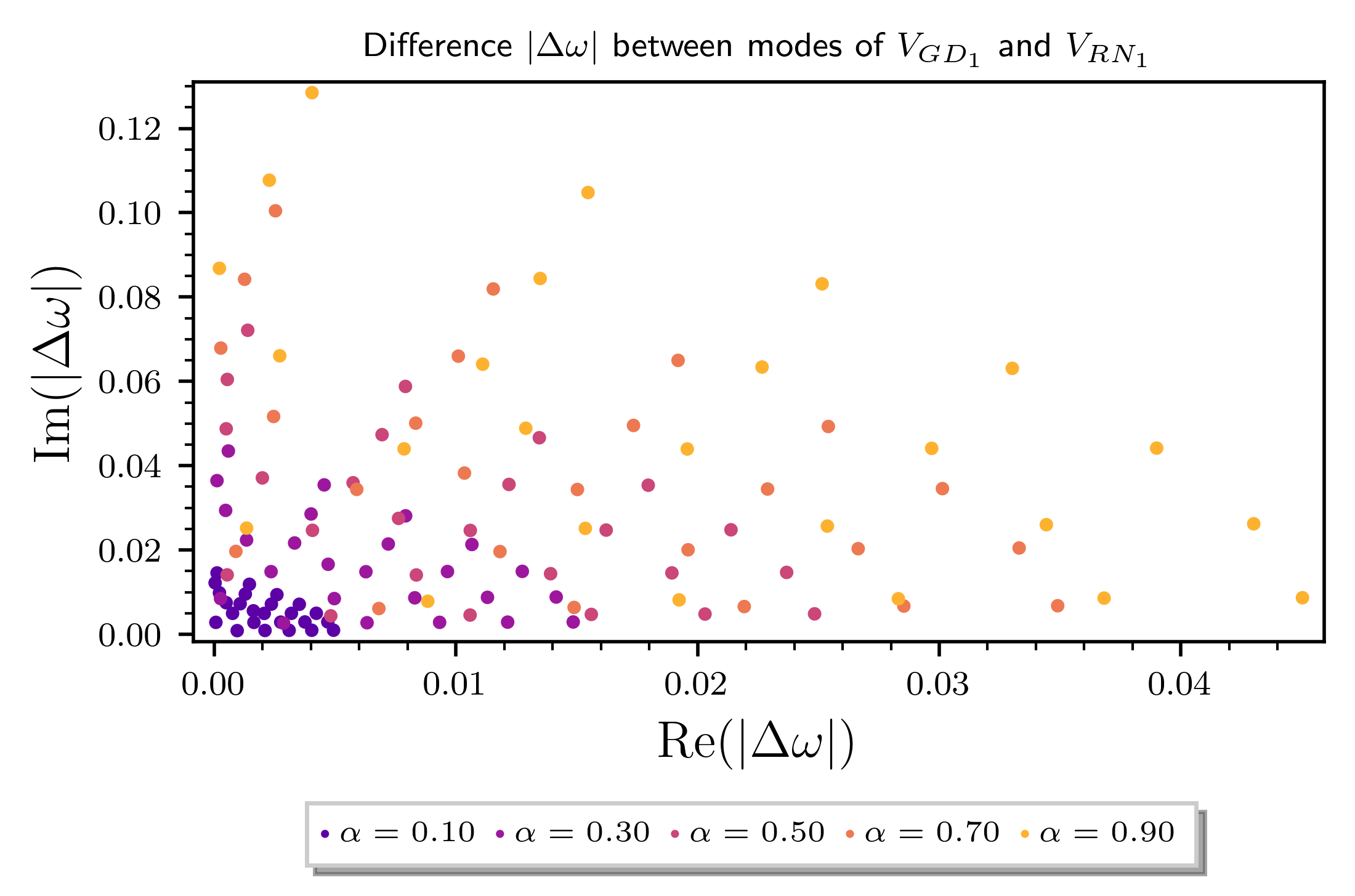}
 \caption{\footnotesize QNMs difference $(\Delta \omega)$ for the potentials $V_{\mathrm{GD_{1}}}$ and $V_{\mathrm{RN_{1}}}$ for a range of values of $\alpha$. It corresponds to the data shown in Table \ref{Delomega1}.}\label{plotDV_1}
\end{figure}

\begin{figure}[ht]
 \centering
\includegraphics[width=0.9\textwidth]{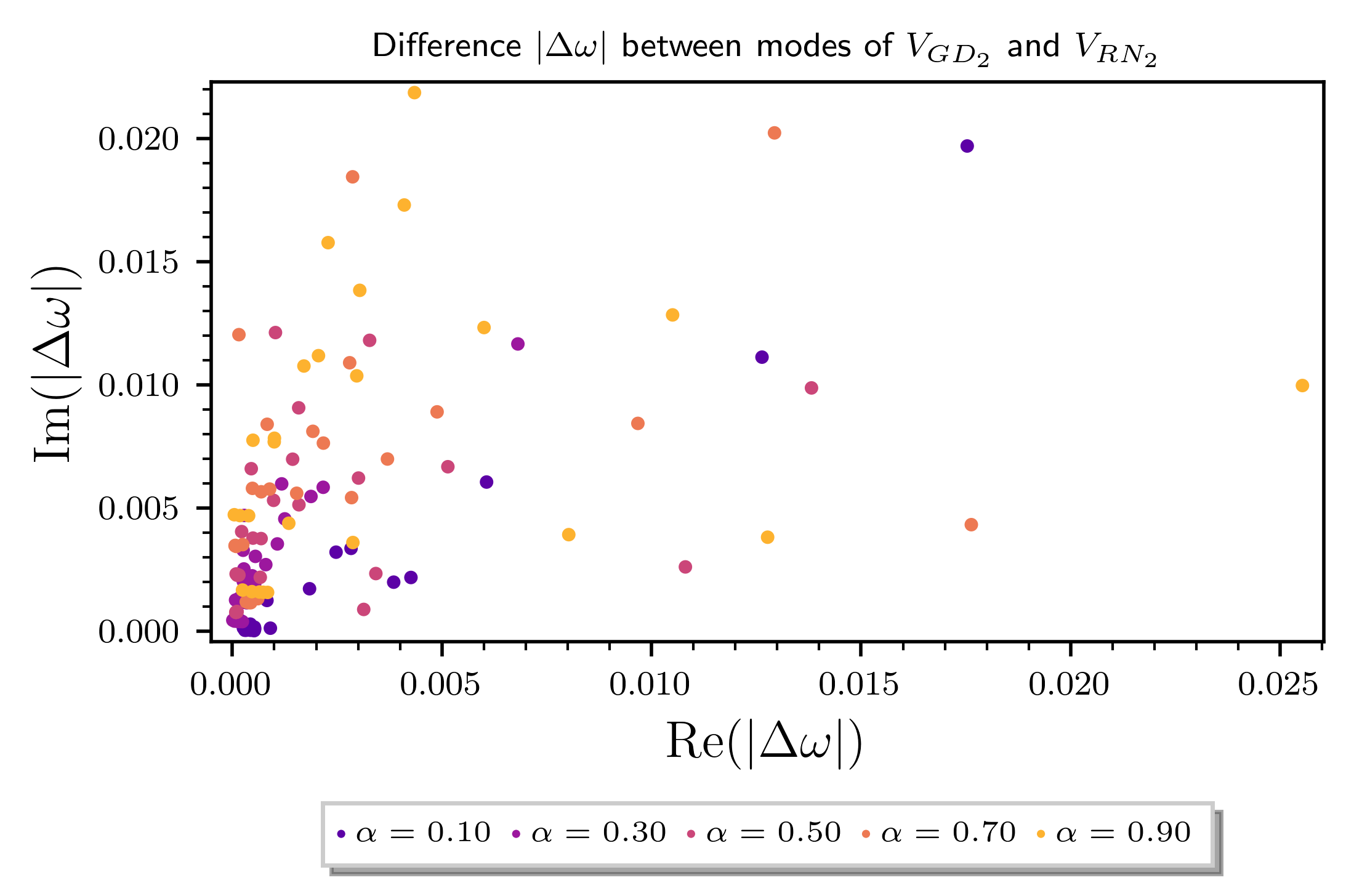}
 \caption{\footnotesize QNMs difference $(\Delta \omega)$ for the potentials $V_{\mathrm{GD_{2}}}$ and $V_{\mathrm{RN_{2}}}$ for a range of values of $\alpha$. It corresponds to the data shown in Table \ref{Delomega2}.}\label{plotDV_2}
\end{figure}

\begin{figure}[ht]
 \centering
\includegraphics[width=0.9\textwidth]{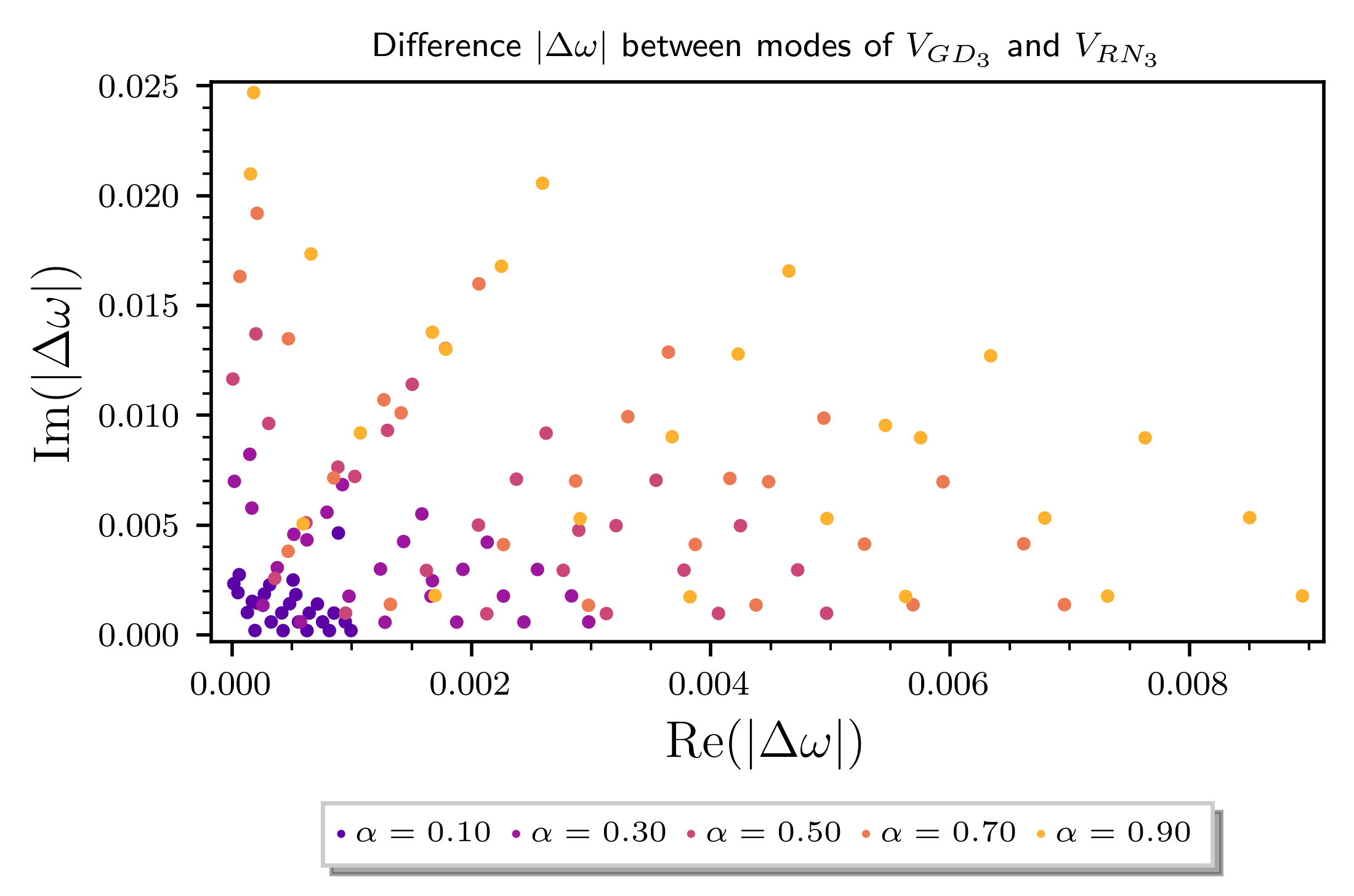}
 \caption{\footnotesize QNMs difference $(\Delta \omega)$ for the potentials $V_{\mathrm{GD_{3}}}$ and $V_{\mathrm{RN_{3}}}$ for a range of values of $\alpha$. It corresponds to the data shown in Table \ref{Delomega3}.}\label{plotDV_3}
\end{figure}

\begin{table}[ht]
\renewcommand{\arraystretch}{1.2}
\centering
\caption{\footnotesize Absolute values of \(\Delta\text{Im}(\omega)\) for the pair $\{V_{\mathrm{GD_{1}}}, V_{\mathrm{RN}_{1}}\}$, \(n_{0}=0\) and $\alpha\in \{0.1, 0.3, 0.5, 0.7, 0.9\}$.}
\label{tableFundModeSelected}
\resizebox{0.7\textwidth}{!}{
\begin{tabular}{||c||c|c|c|c|c||}
\hline\hline
\multirow{1}*{\(n\)} & \multicolumn{1}{c|}{\(\alpha = 0.1\)} & \multicolumn{1}{c|}{\(\alpha = 0.3\)} & \multicolumn{1}{c|}{\(\alpha = 0.5\)} & \multicolumn{1}{c|}{\(\alpha = 0.7\)} & \multicolumn{1}{c||}{\(\alpha = 0.9\)} \\
\thickhline
\;2 \;&\; \(8.83 \times 10^{-4}\) \; & \; \(2.64 \times 10^{-3}\) \; & \; \(4.39 \times 10^{-3}\) \; & \; \(6.13 \times 10^{-3}\) \; & \; \(7.85 \times 10^{-3}\)\; \\
\hline
\;3 \;&\; \(9.22 \times 10^{-4}\) \; & \; \(2.76 \times 10^{-3}\) \; & \; \(4.58 \times 10^{-3}\) \; & \; \(6.39 \times 10^{-3}\) \; & \; \(8.18 \times 10^{-3}\)\; \\
\hline
\;4 \;&\; \(9.53 \times 10^{-4}\) \; & \; \(2.85 \times 10^{-3}\) \; & \; \(4.73 \times 10^{-3}\) \; & \; \(6.60 \times 10^{-3}\) \; & \; \(8.44 \times 10^{-3}\)\; \\
\hline
\;5 \;&\; \(9.71 \times 10^{-4}\) \; & \; \(2.90 \times 10^{-3}\) \; & \; \(4.82 \times 10^{-3}\) \; & \; \(6.72 \times 10^{-3}\) \; & \; \(8.60 \times 10^{-3}\)\; \\
\hline
\;6 \;&\; \(9.81 \times 10^{-4}\) \; & \; \(2.93 \times 10^{-3}\) \; & \; \(4.87 \times 10^{-3}\) \; & \; \(6.79 \times 10^{-3}\) \; & \; \(8.69 \times 10^{-3}\)\; \\
\hline\hline
\end{tabular}}
\label{tableFundMode}
\end{table}

In Table \ref{tableFundMode}  we present the difference of the imaginary parts of the fundamental modes,  for $\alpha\in \{0.1, 0.3, 0.5, 0.7, 0.9\}$. The values from the two potentials deviate approximately ten times more for $\alpha=1.0$, in comparison with $\alpha=0.1$, which hints at the influence of the primary hair in distinguishing the GD potential from the RN potential. However, to ratify these results, we need to know the magnitude of the error of the WKB method at the sixth-order for this potential. This can be achieved by implementing the WKB method to one order higher and one order lower, such that: 
\begin{equation}
    \delta_{k} = \frac{1}{2}\left(\omega _{k+1} - \omega_{k-1}\right), 
    \label{deltak}
\end{equation}
where $k$ is the order of the WKB method, $\delta_{k}$ is the error correspondent to that order, and $\omega_{k+1}$ and $\omega_{k-1}$ are the complex frequencies obtained by the WKB method in the $k+1$ and $k-1$ orders, respectively, as proposed and shown to be a good estimation in \cite{Konoplya:2019hlu}. Setting $k=6$, Eq. \eqref{deltak} becomes:
\begin{equation}
    \delta_{6} = \frac{1}{2}\left(\omega_{7} - \omega_{5}\right).
    \label{delta6}
\end{equation}

In Table \ref{tabledeltak}, we present the error obtained from Eq. \eqref{delta6} for the imaginary parts of the fundamental modes of the GD potential $V_{\mathrm{GD}_{1}}$ at sixth-order in the WKB method. Comparing Tables \ref{tableFundMode} and \ref{tabledeltak}, we see that $\Delta\omega$ is between two and four orders of magnitude higher than $\delta_{6}$, which shows that a comparative spectroscopic analysis of QNMs is indeed capable of distinguishing the GD hairy black hole solutions here presented from possible no-hair solutions, at least in the context where such solutions possess the same values for $\{r_{+}, Q^2\}$.

\begin{table}[ht]
\renewcommand{\arraystretch}{1.2}
\centering
\caption{\footnotesize Imaginary parts of \(\delta _6\) for $V_{\mathrm{GD_{1}}}$, \(n_0 = 0\), and $\alpha\in \{0.1, 0.3, 0.5, 0.7, 0.9\}$.}
\label{tabledeltak}
\resizebox{0.65\textwidth}{!}{
\begin{tabular}{||c||c|c|c|c|c||}
\hline\hline
\(n\) & \(\alpha = 0.1\) & \(\alpha = 0.3\) & \(\alpha = 0.5\) & \(\alpha = 0.7\) & \(\alpha = 0.9\) \\ \thickhline
\;2 \;&\;  \(6.05 \times 10^{-5}\)\color{black} \;&\;  \(6.67 \times 10^{-5}\)\color{black} \;&\;  \(7.57 \times 10^{-5}\)\color{black} \;&\;  \(8.47 \times 10^{-5}\)\color{black} \;&\;  \(9.54 \times 10^{-5}\)\color{black} \\ \hline
\;3 \;&\;  \(4.37 \times 10^{-8}\)\color{black} \;&\;  \(1.96 \times 10^{-7}\)\color{black} \;&\;  \(9.84 \times 10^{-8}\)\color{black} \;&\;  \(2.39 \times 10^{-7}\)\color{black} \;&\;  \(1.80 \times 10^{-7}\)\color{black} \\ \hline
\;4 \;&\;  \(6.48 \times 10^{-8}\)\color{black} \;&\;  \(1.00 \times 10^{-7}\)\color{black} \;&\;  \(1.28 \times 10^{-7}\)\color{black} \;&\;  \(1.62 \times 10^{-7}\)\color{black} \;&\;  \(2.05 \times 10^{-7}\)\color{black} \\ \hline
\;5 \;&\;  \(2.66 \times 10^{-8}\)\color{black} \;&\;  \(2.82 \times 10^{-8}\)\color{black} \;&\;  \(3.78 \times 10^{-8}\)\color{black} \;&\;  \(4.29 \times 10^{-8}\)\color{black} \;&\;  \(4.10 \times 10^{-8}\)\color{black} \\ \hline
\;6 \;&\;  \(9.90 \times 10^{-9}\)\color{black} \;&\;  \(1.14 \times 10^{-8}\)\color{black} \;&\;  \(1.63 \times 10^{-8}\)\color{black} \;&\;  \(1.03 \times 10^{-8}\)\color{black} \;&\;  \(1.71 \times 10^{-8}\)\color{black} \\ \hline\hline
\end{tabular}}
\end{table}
\section{Conclusions}
\label{secVI}
The results presented here attest to the promising nature of the GD method in decoupling sources of gravity and, in the process, producing charges capable of generating primary hair, and the relevance of studying the QNMs of these hairy solutions. {\color{black}{Our main aim in this work was to investigate whether it is theoretically possible to distinguish between solutions arising from the gravitational decoupling method and the Reissner-Nordström solutions with the same pair $\{r_+, Q^2\}$, based on the QNM spectra of tensor perturbations. Our results indicate that such a distinction is indeed feasible, as the differences in the quasi-normal frequencies between the two classes of solutions exceed the estimated error of the WKB method. Thus, confirming $\Delta\omega$ as a hair signature in an observable quantity. However, within the allowed range of the coupling constant $\alpha$, these differences remain below the sensitivity threshold of current gravitational wave detectors \cite{Capote:2024rmo,Abbott:2016xvh}. Nevertheless, this theoretical distinction could become observable with future generations of detectors.}}


\subsection*{\textbf{Acknowledgements}}
VFG thanks CAPES (Grant No.~001). RTC thanks the National Council for Scientific and Technological Development - CNPq (Grant No. 401567/2023-0), for partial financial support. The work of RdR is supported by The S\~ao Paulo Research Foundation (FAPESP) 
(Grants No. 2021/01089-1 and No. 2024/05676-7) and CNPq (Grants No. 303742/2023-2 and No. 401567/2023-0).
\vfill
\newpage

\appendix
\section{The Odd GD potentials}
\label{AOddGD}
Here we present the GD potentials  $V_{\mathrm{GD_{1}}}$, $V_{\mathrm{GD_{2}}}$ and $V_{\mathrm{GD_{3}}}$ in their explicit forms, obtained by substituting the GD metric functions $f_{\mathrm{GD_{1}}}$, $f_{\mathrm{GD_{2}}}$ and $f_{\mathrm{GD_{3}}}$ into Eq. \eqref{RNRW}, respectively.
\begin{align}
V_{\mathrm{GD_{1}}}\left(r\right) 
    &= \,\,\,\Bigg\{ {\left({1} - \frac{2}{r}\right)}\left(\frac{n(n+1)}{r^2}  - \frac{6}{r^{3}} \right)\Bigg\}\,\,  - \label{vgd1}\\[0.3cm]
    & - \alpha\,\Bigg\{ {\left[\frac{1}{2r} \!+\! \frac{{n(n\!+\!1) \!-\!1}}{r^{3}} \!+\! \frac{1}{r^{2}}  \!-\! \frac{12}{r^{4}}\right]} e^{-r}\!+\! {\left[\displaystyle {n(n+1)} {\left(\frac{1}{r^{3}} \!-\! \frac{4}{r^{4}}\right)} + \frac{3}{r^{3}} - \frac{28}{r^{4}} + \frac{56}{r^{5}}\right]} e^{-2}\Bigg\}\,\, + \nonumber \\[0.3cm]
    & + \alpha^2\,\Bigg\{\displaystyle {\left(\frac{3}{r^{4}} - \frac{28}{r^{5}} + \frac{64}{r^{6}}\right)} e^{-4} + \frac{1}{2} {\left(\frac{1}{r^{2}} \!+\! \frac{4}{r^{3}} \!+\! \frac{6}{r^{4}}\right)} e^{-2r} \!+\! \frac{1}{2} {\left(\frac{1}{r^{2}} \!-\! \frac{4}{r^{4}} - \frac{56}{r^{5}}\right)} e^{-r - 2}\Bigg\}\,, \nonumber
    \\[1.2cm]
V_{\mathrm{GD_{2}}}\left(r\right) 
    &= \,\,\,\Bigg\{ {\left({1} - \frac{2}{r}\right)}\left(\frac{n(n+1)}{r^2}  - \frac{6}{r^{3}} \right)\Bigg\}\,\, -   \label{vgd2}\\[0.3cm]
    & - \alpha\Bigg\{{\left[\frac{1}{2r} + \frac{\, {n(n+1) -1}}{r^{3}} + \frac{1}{r^{2}}  - \frac{12}{r^{4}}\right]} e^{-r}\!+\!\ell\left[n(n+1)\left(\frac{1}{r^3} \!-\! \frac{2}{r^4} \right) \!+\! \frac{3}{r^3} \!-\! \frac{20}{r^4} + \frac{28}{r^5}\right]\Bigg\}\,\, \nonumber \\[0.3cm]
    & + \alpha^2\Bigg\{\ell\left[{\left(\frac{{n(n+1)}}{r^{4}} + \frac{4}{r^{4}} - \frac{14}{r^{5}}\right)} e^{-\alpha\ell} + \frac{1}{2} \, {\left(\frac{1}{r^{2}} + \frac{2}{r^{3}} + \frac{4}{r^{4}} - \frac{28}{r^{5}}\right)} e^{-r}\right]\,\,+ \nonumber\\[0.3cm]
    &\quad +\ell^2\left(\displaystyle \frac{3}{r^{4}} - \frac{14}{r^{5}} + \frac{16}{r^{6}}\right) + \frac{1}{2} \, {\left(\frac{1}{r^{2}} + \frac{4}{r^{3}} + \frac{6}{r^{4}}\right)} e^{-2r}\Bigg\}\,\, - \nonumber \\[0.3cm]
    & - \alpha^3\Bigg\{\ell^{2}\left(\frac{7}{r^5} - \frac{16}{r^6}\right)e^{-\alpha\ell} + \ell\left(\frac{1}{2r^3} + \frac{4}{r^4} + \frac{14}{r^5}\right)e^{-\alpha\ell-r}\Bigg\}+\alpha^4\left(\frac{4\ell^2e^{-2\alpha\ell}}{r^6}\right)\,\,, \nonumber
    \\[1.2cm]
V_{\mathrm{GD_{3}}}\left(r\right) 
    &= \,\,\,\Bigg\{ {\left({1} - \frac{2}{r}\right)}\left(\frac{n(n+1)}{r^2}  - \frac{6}{r^{3}} \right)\Bigg\}\,\, - \label{vgd3}  \\[0.3cm] 
    & - \alpha\Bigg\{{\left[\frac{1}{2r} + \frac{\, {n(n+1) -1}}{r^{3}} + \frac{1}{r^{2}}  - \frac{12}{r^{4}}\right]} e^{-r}\,\, +\nonumber\\[0.3cm]
    &\quad + \displaystyle \ell {\left(\frac{{n(n+1)}}{r^{3}} + \frac{3}{r^{3}} - \frac{12}{r^{4}}\right)} - 2 \, {\left(\frac{{n(n+1)}}{r^{4}} + \frac{4}{r^{4}} - \frac{14}{r^{5}}\right)} e^{\left(-\alpha\ell - 2\right)}\Bigg\}\,\, + \nonumber \\[0.3cm]
    & + \alpha^2\Bigg\{\ell\left[{\left(\frac{{n(n+1)} }{r^{4}} + \frac{4 \,}{r^{4}} - \frac{28 \, }{r^{5}}\right)} e^{\left(-\alpha \ell- 2\right)} + {\left(\frac{1}{2r^{2}} + \frac{2}{r^{3}} + \frac{6}{r^{4}}\right)} e^{-r}\right]\,\, + \nonumber\\[0.3cm]  
    &\quad + {\left(\frac{1}{2r^{2}} + \frac{2}{r^{3}} + \frac{3}{r^{4}}\right)} e^{\left(-2 \, r\right)} -{\left(\frac{1}{r^{3}} + \frac{4}{r^{4}} + \frac{14}{r^{5}}\right)} e^{\left(-\alpha \ell- r - 2\right)} +\frac{16e^{-2 \, \alpha \ell- 4}}{r^{6}} + \frac{3 \, \ell^{2}}{r^{4}} \Bigg\}\,\,  \nonumber \\[0.3cm]
    & - \alpha^{3}\Bigg\{\ell\left[{\left(\frac{1}{2r^{3}} + \frac{2}{r^{4}}  + \frac{7}{r^{5}}\right)} e^{\left(-\alpha\ell - r - 2\right)}- \frac{16 e^{-2 \, \alpha\ell - 4}}{r^{6}}\right] + \frac{7\ell^{2} e^{\left(-\alpha\ell - 2\right)}}{r^{5}} \Bigg\}+\alpha^{4}\left(\frac{4 \, \ell^{2} e^{-2 \, \alpha\ell - 4}}{r^{6}}\right). \nonumber
\end{align}

\label{appendixB}
\newpage
\section{QNMs from GD Potentials}
We now display the real and imaginary parts of the complex frequencies that compose the QNMs obtained through the sixth-order WKB method applied to the GD potentials $V_{\mathrm{GD_{1}}}$, $V_{\mathrm{GD_{2}}}$ and $V_{\mathrm{GD_{3}}}$, whose explicit forms can be found in Appendix \ref{AOddGD}.

\begin{table}[H]
  \centering  
  \textbf{QNMs from \(\boldsymbol{V_{\mathrm{GD}_{1}}}\)}\\[8pt]
  \fontsize{10}{12}\selectfont
  \begin{adjustbox}{width=\textwidth}
    \setlength{\tabcolsep}{5pt}
    \begin{tabular}{||c||c||cc|cc|cc|cc|cc||}
      \hline\hline
      \multirow{2}{*}{$n$} & \multirow{2}{*}{$n_0$}
        & \multicolumn{2}{c|}{$\alpha=0.1$}
        & \multicolumn{2}{c|}{$\alpha=0.3$}
        & \multicolumn{2}{c|}{$\alpha=0.5$}
        & \multicolumn{2}{c|}{$\alpha=0.7$}
        & \multicolumn{2}{c||}{$\alpha=0.9$} \\
      \cline{3-12}
      & & Re\,$\omega$ & Im\,$\omega$
        & Re\,$\omega$ & Im\,$\omega$
        & Re\,$\omega$ & Im\,$\omega$
        & Re\,$\omega$ & Im\,$\omega$
        & Re\,$\omega$ & Im\,$\omega$ \\
      \thickhline
      \multirow{3}{*}{2}
        & 0 & 0.37303 & -0.08886 & 0.37184 & -0.08880 & 0.37065 & -0.08875 & 0.36946 & -0.08870 & 0.36828 & -0.08865 \\
        & 1 & 0.34553 & -0.27343 & 0.34398 & -0.27334 & 0.34241 & -0.27326 & 0.34083 & -0.27321 & 0.33923 & -0.27317 \\
        & 2 & 0.29739 & -0.47774 & 0.29509 & -0.47811 & 0.29276 & -0.47850 & 0.29038 & -0.47891 & 0.28797 & -0.47937 \\
      \hline
      \multirow{4}{*}{3}
        & 0 & 0.59894 & -0.09267 & 0.59794 & -0.09260 & 0.59694 & -0.09253 & 0.59594 & -0.09247 & 0.59494 & -0.09240 \\
        & 1 & 0.58208 & -0.28119 & 0.58094 & -0.28099 & 0.57981 & -0.28078 & 0.57868 & -0.28058 & 0.57754 & -0.28038 \\
        & 2 & 0.55088 & -0.47888 & 0.54945 & -0.47856 & 0.54801 & -0.47823 & 0.54656 & -0.47791 & 0.54511 & -0.47758 \\
        & 3 & 0.51012 & -0.69028 & 0.50816 & -0.68990 & 0.50617 & -0.68953 & 0.50417 & -0.68916 & 0.50215 & -0.68878 \\
      \hline
      \multirow{5}{*}{4}
        & 0 & 0.80871 & -0.09414 & 0.80777 & -0.09409 & 0.80683 & -0.09405 & 0.80589 & -0.09399 & 0.80495 & -0.09395 \\
        & 1 & 0.79611 & -0.28426 & 0.79506 & -0.28411 & 0.79401 & -0.28396 & 0.79296 & -0.28382 & 0.79191 & -0.28367 \\
        & 2 & 0.77205 & -0.47977 & 0.77076 & -0.47951 & 0.76946 & -0.47926 & 0.76816 & -0.47901 & 0.76686 & -0.47876 \\
        & 3 & 0.73881 & -0.68372 & 0.73708 & -0.68337 & 0.73534 & -0.68302 & 0.73360 & -0.68268 & 0.73185 & -0.68235 \\
        & 4 & 0.69944 & -0.89827 & 0.69702 & -0.89789 & 0.69458 & -0.89753 & 0.69213 & -0.89719 & 0.68965 & -0.89686 \\
      \hline
      \multirow{6}{*}{5}
        & 0 & 1.01183 & -0.09485 & 1.01090 & -0.09482 & 1.00996 & -0.09478 & 1.00903 & -0.09475 & 1.00810 & -0.09472 \\
        & 1 & 1.00171 & -0.28576 & 1.00068 & -0.28565 & 0.99965 & -0.28555 & 0.99863 & -0.28544 & 0.99760 & -0.28534 \\
        & 2 & 0.98208 & -0.48023 & 0.98085 & -0.48004 & 0.97962 & -0.47985 & 0.97838 & -0.47967 & 0.97715 & -0.47949 \\
        & 3 & 0.95417 & -0.68040 & 0.95259 & -0.68013 & 0.95101 & -0.67986 & 0.94942 & -0.67960 & 0.94783 & -0.67935 \\
        & 4 & 0.91975 & -0.88802 & 0.91764 & -0.88769 & 0.91551 & -0.88737 & 0.91337 & -0.88707 & 0.91122 & -0.88679 \\
        & 5 & 0.88093 & -1.10429 & 0.87801 & -1.10399 & 0.87508 & -1.10372 & 0.87213 & -1.10349 & 0.86916 & -1.10328 \\
      \hline
      \multirow{7}{*}{6}
        & 0 & 1.21153 & -0.09525 & 1.21058 & -0.09523 & 1.20962 & -0.09520 & 1.20867 & -0.09517 & 1.20771 & -0.09515 \\
        & 1 & 1.20306 & -0.28661 & 1.20202 & -0.28653 & 1.20098 & -0.28645 & 1.19994 & -0.28637 & 1.19891 & -0.28629 \\
        & 2 & 1.18647 & -0.48049 & 1.18526 & -0.48034 & 1.18405 & -0.48020 & 1.18283 & -0.48006 & 1.18162 & -0.47993 \\
        & 3 & 1.16251 & -0.67847 & 1.16101 & -0.67826 & 1.15951 & -0.67805 & 1.15801 & -0.67785 & 1.15650 & -0.67766 \\
        & 4 & 1.13226 & -0.88194 & 1.13033 & -0.88167 & 1.12839 & -0.88092 & 1.12645 & -0.88081 & 1.12450 & -0.88092 \\
        & 5 & 1.09710 & -1.09197 & 1.09454 & -1.09168 & 1.09198 & -1.09095 & 1.08940 & -1.09117 & 1.08681 & -1.09095 \\
        & 6 & 1.05853 & -1.30933 & 1.05512 & -1.30912 & 1.05169 & -1.30872 & 1.04823 & -1.30881 & 1.04476 & -1.30872 \\
      \hline\hline
    \end{tabular}
  \end{adjustbox}
  \caption{\footnotesize Real and imaginary parts of the QNMs of \(V_{\mathrm{GD}_{1}}\) for \(\alpha = 0.1,0.3,0.5,0.7,0.9\).}
  \label{qnmVGD1}
\end{table}

\begin{table}[H]
  \centering
  \textbf{QNMs from \(\boldsymbol{V_{\mathrm{GD}_{2}}}\)}\\[8pt]
  \fontsize{10}{12}\selectfont
  \begin{adjustbox}{width=\textwidth}
    \setlength{\tabcolsep}{5pt}
    \begin{tabular}{||c||c||cc|cc|cc|cc|cc||}
      \hline\hline
      \multirow{2}{*}{$n$} & \multirow{2}{*}{$n_0$}
        & \multicolumn{2}{c|}{$\alpha=0.1$}
        & \multicolumn{2}{c|}{$\alpha=0.3$}
        & \multicolumn{2}{c|}{$\alpha=0.5$}
        & \multicolumn{2}{c|}{$\alpha=0.7$}
        & \multicolumn{2}{c||}{$\alpha=0.9$} \\
      \cline{3-12}
      & & Re\,$\omega$ & Im\,$\omega$
        & Re\,$\omega$ & Im\,$\omega$
        & Re\,$\omega$ & Im\,$\omega$
        & Re\,$\omega$ & Im\,$\omega$
        & Re\,$\omega$ & Im\,$\omega$ \\
      \thickhline
      \multirow{3}{*}{2}
        & 0 & 0.19256 & -0.03387 & 0.19280 & -0.03397 & 0.19315 & -0.03407 & 0.19324 & -0.03420 & 0.19369 & -0.03428 \\
        & 1 & 0.18480 & -0.10305 & 0.18487 & -0.10348 & 0.18546 & -0.10290 & 0.18441 & -0.10468 & 0.18661 & -0.10393 \\
        & 2 & 0.17014 & -0.17662 & 0.16950 & -0.17808 & 0.17248 & -0.17453 & 0.16552 & -0.18401 & 0.17432 & -0.17555 \\
      \hline
      \multirow{4}{*}{3}
        & 0 & 0.30756 & -0.03533 & 0.30804 & -0.03540 & 0.30778 & -0.03537 & 0.30904 & -0.03556 & 0.30954 & -0.03564 \\
        & 1 & 0.30274 & -0.10647 & 0.30314 & -0.10674 & 0.30268 & -0.10669 & 0.30443 & -0.10713 & 0.30496 & -0.10737 \\
        & 2 & 0.29380 & -0.17873 & 0.29366 & -0.17957 & 0.29227 & -0.18004 & 0.29559 & -0.18048 & 0.29668 & -0.18005 \\
        & 3 & 0.28217 & -0.25163 & 0.28022 & -0.25459 & 0.27610 & -0.25774 & 0.28589 & -0.25198 & 0.28649 & -0.25268 \\
      \hline
      \multirow{5}{*}{4}
        & 0 & 0.41428 & -0.03585 & 0.41497 & -0.03592 & 0.41462 & -0.03588 & 0.41635 & -0.03605 & 0.41704 & -0.03613 \\
        & 1 & 0.41049 & -0.10787 & 0.41124 & -0.10807 & 0.41085 & -0.10797 & 0.41275 & -0.10846 & 0.41340 & -0.10870 \\
        & 2 & 0.40290 & -0.18090 & 0.40382 & -0.18118 & 0.40326 & -0.18108 & 0.40571 & -0.18176 & 0.40599 & -0.18235 \\
        & 3 & 0.39139 & -0.25582 & 0.39272 & -0.25604 & 0.39167 & -0.25617 & 0.39558 & -0.25637 & 0.39447 & -0.25821 \\
        & 4 & 0.37572 & -0.33398 & 0.37791 & -0.33364 & 0.37572 & -0.33474 & 0.39609 & -0.33077 & 0.37813 & -0.33830 \\
      \hline
      \multirow{6}{*}{5}
        & 0 & 0.51772 & -0.03609 & 0.51859 & -0.03615 & 0.51815 & -0.03612 & 0.52035 & -0.03628 & 0.52124 & -0.03635 \\
        & 1 & 0.51470 & -0.10848 & 0.51560 & -0.10867 & 0.51514 & -0.10858 & 0.51745 & -0.10905 & 0.51839 & -0.10925 \\
        & 2 & 0.50870 & -0.18148 & 0.50965 & -0.18180 & 0.50909 & -0.18166 & 0.51177 & -0.18239 & 0.51185 & -0.18275 \\
        & 3 & 0.49980 & -0.25550 & 0.50078 & -0.25599 & 0.49995 & -0.25591 & 0.50352 & -0.25662 & 0.50499 & -0.25691 \\
        & 4 & 0.48811 & -0.33097 & 0.48899 & -0.33178 & 0.48753 & -0.33205 & 0.49311 & -0.33185 & 0.49549 & -0.33173 \\
        & 5 & 0.47375 & -0.40835 & 0.47428 & -0.40983 & 0.47150 & -0.41122 & 0.48110 & -0.40790 & 0.48532 & -0.40637 \\
      \hline
      \multirow{7}{*}{6}
        & 0 & 0.61947 & -0.03623 & 0.62053 & -0.03628 & 0.61999 & -0.03625 & 0.62265 & -0.03641 & 0.62372 & -0.03647 \\
        & 1 & 0.61694 & -0.10882 & 0.61803 & -0.10900 & 0.61748 & -0.10891 & 0.62022 & -0.10937 & 0.62131 & -0.10956 \\
        & 2 & 0.61188 & -0.18185 & 0.61307 & -0.18213 & 0.61243 & -0.18200 & 0.61539 & -0.18275 & 0.61652 & -0.18309 \\
        & 3 & 0.60430 & -0.25563 & 0.60572 & -0.25597 & 0.60481 & -0.25589 & 0.60824 & -0.25684 & 0.60936 & -0.25735 \\
        & 4 & 0.59417 & -0.33054 & 0.59609 & -0.33077 & 0.59449 & -0.33101 & 0.59887 & -0.33189 & 0.59984 & -0.33270 \\
        & 5 & 0.58139 & -0.40707 & 0.58432 & -0.40983 & 0.58123 & -0.40805 & 0.58745 & -0.40813 & 0.58791 & -0.40959 \\
        & 6 & 0.56585 & -0.48591 & 0.57060 & -0.48414 & 0.56468 & -0.48803 & 0.57413 & -0.48577 & 0.57350 & -0.48862 \\
      \hline\hline
    \end{tabular}
  \end{adjustbox}
  \caption{\footnotesize Real and imaginary parts of the QNMs of \(V_{\mathrm{GD}_{2}}\) for \(\alpha = 0.1,0.3,0.5,0.7,0.9\).}
  \label{qnmVGD2}
\end{table}

\begin{table}[H]
  \centering
  \textbf{QNMs from \(\boldsymbol{V_{\mathrm{GD}_3}}\)}\\[8pt]
  \fontsize{10}{12}\selectfont
  \begin{adjustbox}{width=\textwidth}
    \setlength{\tabcolsep}{5pt}
    \begin{tabular}{||c||c||cc|cc|cc|cc|cc||}
      \hline\hline
      \multirow{2}{*}{$n$} & \multirow{2}{*}{$n_0$}
        & \multicolumn{2}{c|}{$\alpha=0.1$}
        & \multicolumn{2}{c|}{$\alpha=0.3$}
        & \multicolumn{2}{c|}{$\alpha=0.5$}
        & \multicolumn{2}{c|}{$\alpha=0.7$}
        & \multicolumn{2}{c||}{$\alpha=0.9$} \\
      \cline{3-12}
      & & Re\,$\omega$ & Im\,$\omega$
        & Re\,$\omega$ & Im\,$\omega$
        & Re\,$\omega$ & Im\,$\omega$
        & Re\,$\omega$ & Im\,$\omega$
        & Re\,$\omega$ & Im\,$\omega$ \\
      \thickhline
      \multirow{3}{*}{2}
        & 0 & 0.24914 & -0.05938 & 0.24927 & -0.05963 & 0.24921 & -0.05951 & 0.24952 & -0.06013 & 0.24964 & -0.06038 \\
        & 1 & 0.23076 & -0.18269 & 0.23054 & -0.18343 & 0.23065 & -0.18305 & 0.23006 & -0.18494 & 0.22980 & -0.18572 \\
        & 2 & 0.19848 & -0.31903 & 0.19738 & -0.32039 & 0.19794 & -0.31971 & 0.19506 & -0.32328 & 0.19383 & -0.32482 \\
      \hline
      \multirow{4}{*}{3}
        & 0 & 0.39984 & -0.06191 & 0.40027 & -0.06214 & 0.40006 & -0.06203 & 0.40112 & -0.06260 & 0.40154 & -0.06283 \\
        & 1 & 0.38857 & -0.18787 & 0.38887 & -0.18856 & 0.38872 & -0.18822 & 0.38943 & -0.18995 & 0.38971 & -0.19065 \\
        & 2 & 0.36774 & -0.31996 & 0.36775 & -0.32116 & 0.36774 & -0.32056 & 0.36774 & -0.32357 & 0.36772 & -0.32478 \\
        & 3 & 0.34053 & -0.46121 & 0.34010 & -0.46300 & 0.34032 & -0.46210 & 0.33919 & -0.46661 & 0.33870 & -0.46843 \\
      \hline
      \multirow{5}{*}{4}
        & 0 & 0.53979 & -0.06289 & 0.54046 & -0.06312 & 0.54012 & -0.06300 & 0.54179 & -0.06357 & 0.54246 & -0.06380 \\
        & 1 & 0.53138 & -0.18990 & 0.53195 & -0.19059 & 0.53166 & -0.19024 & 0.53309 & -0.19197 & 0.53365 & -0.19267 \\
        & 2 & 0.51532 & -0.32051 & 0.51571 & -0.32168 & 0.51552 & -0.32109 & 0.51645 & -0.32402 & 0.51627 & -0.32343 \\
        & 3 & 0.49317 & -0.45677 & 0.49327 & -0.45843 & 0.49322 & -0.45760 & 0.49342 & -0.46178 & 0.49348 & -0.46347 \\
        & 4 & 0.46697 & -0.60007 & 0.46669 & -0.60228 & 0.46683 & -0.60118 & 0.46606 & -0.60671 & 0.46570 & -0.60893 \\
      \hline
      \multirow{6}{*}{5}
        & 0 & 0.67531 & -0.06336 & 0.67620 & -0.06359 & 0.67575 & -0.06347 & 0.67798 & -0.06405 & 0.67887 & -0.06428 \\
        & 1 & 0.66856 & -0.19089 & 0.66937 & -0.19158 & 0.66896 & -0.19123 & 0.67099 & -0.19296 & 0.67180 & -0.19366 \\
        & 2 & 0.65546 & -0.32079 & 0.65613 & -0.32195 & 0.65580 & -0.32137 & 0.65744 & -0.32428 & 0.65678 & -0.32545 \\
        & 3 & 0.63686 & -0.45451 & 0.63730 & -0.45615 & 0.63708 & -0.45533 & 0.63815 & -0.45944 & 0.63773 & -0.46110 \\
        & 4 & 0.61395 & -0.59319 & 0.61410 & -0.59532 & 0.61403 & -0.59425 & 0.61432 & -0.59960 & 0.61440 & -0.60175 \\
        & 5 & 0.58815 & -0.73762 & 0.58793 & -0.74025 & 0.58804 & -0.73893 & 0.58738 & -0.74554 & 0.58705 & -0.74819 \\
      \hline
      \multirow{7}{*}{6}
        & 0 & 0.80856 & -0.06362 & 0.80966 & -0.06385 & 0.80911 & -0.06374 & 0.81187 & -0.06432 & 0.81297 & -0.06455 \\
        & 1 & 0.80290 & -0.19144 & 0.80394 & -0.19213 & 0.80342 & -0.19179 & 0.80602 & -0.19353 & 0.80705 & -0.19423 \\
        & 2 & 0.79184 & -0.32095 & 0.79275 & -0.32211 & 0.79230 & -0.32153 & 0.79456 & -0.32444 & 0.79546 & -0.32561 \\
        & 3 & 0.77587 & -0.45320 & 0.77659 & -0.45483 & 0.77623 & -0.45401 & 0.77801 & -0.45811 & 0.77870 & -0.45977 \\
        & 4 & 0.75573 & -0.58910 & 0.75620 & -0.59121 & 0.75597 & -0.59016 & 0.75710 & -0.59546 & 0.75752 & -0.59759 \\
        & 5 & 0.73234 & -0.72938 & 0.73251 & -0.73197 & 0.73243 & -0.73067 & 0.73275 & -0.73718 & 0.73284 & -0.73980 \\
        & 6 & 0.70673 & -0.87451 & 0.70653 & -0.87758 & 0.70664 & -0.87758 & 0.70600 & -0.88376 & 0.70566 & -0.88686 \\
      \hline\hline
    \end{tabular}
  \end{adjustbox}
  \caption{\footnotesize Real and imaginary parts of the QNMs of \(V_{\mathrm{GD}_{3}}\) for \(\alpha=0.1,0.3,0.5,0.7,0.9\).}
  \label{qnmVGD3}
\end{table}

\newpage
\section{\texorpdfstring{$\Delta\omega$}{Delta omega} between GD and RN black holes with same \texorpdfstring{$\{r_{+}\,,\,Q^{2}\}$}{\{r+, Q2\}}}
\label{AppendixOmega}
Next we display the obtained values for $\Delta\omega$, computed using Eq. \eqref{DeltaOmega} for each pair $\{V_{\mathrm{GD}_{i}}\,,\, V_{{\mathrm{RN}}_{i}}\}$, $i\in\{1,2,3\}$, which compares the QNMs from the GD potentials with the QNMs from the correspondent RN black holes with same values of outer horizon $r_{+}$ and squared charge $Q^2$.

\begin{table}[H]
  \centering
  \renewcommand{\arraystretch}{1.2} 
  \textbf{QNM difference $\bigl(\Delta\omega\bigr)$ for $\{V_{\mathrm{GD}_1},V_{\mathrm{RN}_1}\}$}\\[10pt]
  \fontsize{12}{15}\selectfont
  \begin{adjustbox}{width=\textwidth}
    \setlength{\tabcolsep}{5pt}
    \begin{tabular}{||c||c||
      cc|cc|cc|cc|cc||}
      \hline\hline
      \multirow{2}{*}{$n$} & \multirow{2}{*}{$n_0$}
        & \multicolumn{2}{c|}{$\alpha=0.1$}
        & \multicolumn{2}{c|}{$\alpha=0.3$}
        & \multicolumn{2}{c|}{$\alpha=0.5$}
        & \multicolumn{2}{c|}{$\alpha=0.7$}
        & \multicolumn{2}{c||}{$\alpha=0.9$} \\
      \cline{3-12}
        &  & $\Delta\Re(\omega)$ & $\Delta\Im(\omega)$
        & $\Delta\Re(\omega)$ & $\Delta\Im(\omega)$
        & $\Delta\Re(\omega)$ & $\Delta\Im(\omega)$
        & $\Delta\Re(\omega)$ & $\Delta\Im(\omega)$
        & $\Delta\Re(\omega)$ & $\Delta\Im(\omega)$ \\
      \thickhline
      \multirow{3}{*}{2}
        & 0 & $9.46\times10^{-4}$ & $8.83\times10^{-4}$
            & $2.87\times10^{-3}$ & $2.64\times10^{-3}$
            & $4.82\times10^{-3}$ & $4.39\times10^{-3}$
            & $6.81\times10^{-3}$ & $6.13\times10^{-3}$
            & $8.83\times10^{-3}$ & $7.85\times10^{-3}$ \\
        & 1 & $6.35\times10^{-5}$ & $2.85\times10^{-3}$
            & $1.33\times10^{-3}$ & $2.24\times10^{-2}$
            & $5.31\times10^{-4}$ & $1.41\times10^{-2}$
            & $8.91\times10^{-4}$ & $1.97\times10^{-2}$
            & $1.33\times10^{-3}$ & $2.52\times10^{-2}$ \\
        & 2 & $1.62\times10^{-3}$ & $5.58\times10^{-3}$
            & $2.35\times10^{-3}$ & $1.49\times10^{-2}$
            & $7.62\times10^{-3}$ & $2.75\times10^{-2}$
            & $1.04\times10^{-2}$ & $3.83\times10^{-2}$
            & $1.29\times10^{-2}$ & $4.89\times10^{-2}$ \\
      \hline
      \multirow{4}{*}{3}
        & 0 & $2.10\times10^{-3}$ & $9.22\times10^{-4}$
            & $6.32\times10^{-3}$ & $2.76\times10^{-3}$
            & $1.06\times10^{-2}$ & $4.58\times10^{-3}$
            & $1.49\times10^{-2}$ & $6.39\times10^{-3}$
            & $1.92\times10^{-2}$ & $8.18\times10^{-3}$ \\
        & 1 & $1.64\times10^{-3}$ & $2.84\times10^{-3}$
            & $4.97\times10^{-3}$ & $8.48\times10^{-3}$
            & $8.36\times10^{-3}$ & $1.41\times10^{-2}$
            & $1.18\times10^{-2}$ & $1.96\times10^{-2}$
            & $1.54\times10^{-2}$ & $2.51\times10^{-2}$ \\
        & 2 & $7.54\times10^{-4}$ & $4.98\times10^{-3}$
            & $2.35\times10^{-3}$ & $1.49\times10^{-2}$
            & $4.06\times10^{-3}$ & $2.47\times10^{-2}$
            & $5.89\times10^{-3}$ & $3.44\times10^{-2}$
            & $7.85\times10^{-3}$ & $4.40\times10^{-2}$ \\
        & 3 & $4.90\times10^{-4}$ & $7.50\times10^{-3}$
            & $1.33\times10^{-3}$ & $2.24\times10^{-2}$
            & $1.99\times10^{-3}$ & $3.71\times10^{-2}$
            & $2.45\times10^{-3}$ & $5.17\times10^{-2}$
            & $2.71\times10^{-3}$ & $6.61\times10^{-2}$ \\
      \hline
      \multirow{5}{*}{4}
        & 0 & $3.09\times10^{-3}$ & $9.53\times10^{-4}$
            & $9.32\times10^{-3}$ & $2.85\times10^{-3}$
            & $1.56\times10^{-2}$ & $4.73\times10^{-3}$
            & $2.19\times10^{-2}$ & $6.60\times10^{-3}$
            & $2.83\times10^{-2}$ & $8.44\times10^{-3}$ \\
        & 1 & $2.75\times10^{-3}$ & $2.90\times10^{-3}$
            & $8.30\times10^{-3}$ & $8.67\times10^{-3}$
            & $1.39\times10^{-2}$ & $1.44\times10^{-2}$
            & $1.96\times10^{-2}$ & $2.01\times10^{-2}$
            & $2.54\times10^{-2}$ & $2.57\times10^{-2}$ \\
        & 2 & $2.06\times10^{-3}$ & $4.97\times10^{-3}$
            & $6.27\times10^{-3}$ & $1.49\times10^{-2}$
            & $1.06\times10^{-2}$ & $2.47\times10^{-2}$
            & $1.50\times10^{-2}$ & $3.44\times10^{-2}$
            & $1.96\times10^{-2}$ & $4.40\times10^{-2}$ \\
        & 3 & $1.07\times10^{-3}$ & $7.26\times10^{-3}$
            & $3.32\times10^{-3}$ & $2.17\times10^{-2}$
            & $5.74\times10^{-3}$ & $3.60\times10^{-2}$
            & $8.33\times10^{-3}$ & $5.01\times10^{-2}$
            & $1.11\times10^{-2}$ & $6.41\times10^{-2}$ \\
        & 4 & $2.12\times10^{-4}$ & $9.85\times10^{-3}$
            & $4.68\times10^{-4}$ & $2.94\times10^{-2}$
            & $4.91\times10^{-4}$ & $4.88\times10^{-2}$
            & $2.69\times10^{-4}$ & $6.79\times10^{-2}$
            & $2.10\times10^{-4}$ & $8.68\times10^{-2}$ \\
      \hline
      \multirow{6}{*}{5}
        & 0 & $4.03\times10^{-3}$ & $9.71\times10^{-4}$
            & $1.21\times10^{-2}$ & $2.90\times10^{-3}$
            & $2.03\times10^{-2}$ & $4.82\times10^{-3}$
            & $2.85\times10^{-2}$ & $6.72\times10^{-3}$
            & $3.68\times10^{-2}$ & $8.60\times10^{-3}$ \\
        & 1 & $3.75\times10^{-3}$ & $2.94\times10^{-3}$
            & $1.13\times10^{-2}$ & $8.78\times10^{-3}$
            & $1.89\times10^{-2}$ & $1.46\times10^{-2}$
            & $2.66\times10^{-2}$ & $2.03\times10^{-2}$
            & $3.44\times10^{-2}$ & $2.60\times10^{-2}$ \\
        & 2 & $3.19\times10^{-3}$ & $4.99\times10^{-3}$
            & $9.65\times10^{-3}$ & $1.49\times10^{-2}$
            & $1.62\times10^{-2}$ & $2.47\times10^{-2}$
            & $2.29\times10^{-2}$ & $3.45\times10^{-2}$
            & $2.97\times10^{-2}$ & $4.41\times10^{-2}$ \\
        & 3 & $2.36\times10^{-3}$ & $7.17\times10^{-3}$
            & $7.20\times10^{-3}$ & $2.14\times10^{-2}$
            & $1.22\times10^{-2}$ & $3.56\times10^{-2}$
            & $1.73\times10^{-2}$ & $4.96\times10^{-2}$
            & $2.27\times10^{-2}$ & $6.34\times10^{-2}$ \\
        & 4 & $1.28\times10^{-3}$ & $9.56\times10^{-3}$
            & $4.00\times10^{-3}$ & $2.86\times10^{-2}$
            & $6.94\times10^{-3}$ & $4.74\times10^{-2}$
            & $1.01\times10^{-2}$ & $6.60\times10^{-2}$
            & $1.35\times10^{-2}$ & $8.44\times10^{-2}$ \\
        & 5 & $3.21\times10^{-5}$ & $1.22\times10^{-2}$
            & $1.10\times10^{-4}$ & $3.65\times10^{-2}$
            & $5.37\times10^{-4}$ & $6.05\times10^{-2}$
            & $1.25\times10^{-3}$ & $8.42\times10^{-2}$
            & $2.28\times10^{-3}$ & $1.08\times10^{-1}$ \\
      \hline
      \multirow{7}{*}{6}
        & 0 & $4.93\times10^{-3}$ & $9.81\times10^{-4}$
            & $1.49\times10^{-2}$ & $2.93\times10^{-3}$
            & $2.48\times10^{-2}$ & $4.87\times10^{-3}$
            & $3.49\times10^{-2}$ & $6.79\times10^{-3}$
            & $4.50\times10^{-2}$ & $8.69\times10^{-3}$ \\
        & 1 & $4.69\times10^{-3}$ & $2.96\times10^{-3}$
            & $1.41\times10^{-2}$ & $8.85\times10^{-3}$
            & $2.37\times10^{-2}$ & $1.47\times10^{-2}$
            & $3.33\times10^{-2}$ & $2.05\times10^{-2}$
            & $4.30\times10^{-2}$ & $2.62\times10^{-2}$ \\
        & 2 & $4.22\times10^{-3}$ & $5.00\times10^{-3}$
            & $1.27\times10^{-2}$ & $1.49\times10^{-2}$
            & $2.14\times10^{-2}$ & $2.48\times10^{-2}$
            & $3.01\times10^{-2}$ & $3.46\times10^{-2}$
            & $3.90\times10^{-2}$ & $4.42\times10^{-2}$ \\
        & 3 & $3.52\times10^{-3}$ & $7.14\times10^{-3}$
            & $1.07\times10^{-2}$ & $2.13\times10^{-2}$
            & $1.80\times10^{-2}$ & $3.54\times10^{-2}$
            & $2.54\times10^{-2}$ & $4.93\times10^{-2}$
            & $3.30\times10^{-2}$ & $6.31\times10^{-2}$ \\
        & 4 & $2.59\times10^{-3}$ & $9.41\times10^{-3}$
            & $7.92\times10^{-3}$ & $2.81\times10^{-2}$
            & $1.34\times10^{-2}$ & $4.66\times10^{-2}$
            & $1.07\times10^{-2}$ & $3.74\times10^{-2}$
            & $2.52\times10^{-2}$ & $8.31\times10^{-2}$ \\
        & 5 & $1.45\times10^{-3}$ & $1.19\times10^{-2}$
            & $4.55\times10^{-3}$ & $3.55\times10^{-2}$
            & $1.73\times10^{-3}$ & $9.60\times10^{-2}$
            & $6.19\times10^{-3}$ & $4.72\times10^{-2}$
            & $1.55\times10^{-2}$ & $1.05\times10^{-1}$ \\
        & 6 & $1.12\times10^{-4}$ & $1.46\times10^{-2}$
            & $5.80\times10^{-4}$ & $4.35\times10^{-2}$
            & $1.38\times10^{-3}$ & $7.21\times10^{-2}$
            & $9.40\times10^{-4}$ & $5.78\times10^{-2}$
            & $4.04\times10^{-3}$ & $1.28\times10^{-1}$ \\
      \hline\hline
    \end{tabular}
  \end{adjustbox}
  \caption{\footnotesize Real and imaginary parts of the difference $\Delta\omega$ between QNMs of potentials $V_{\mathrm{GD}_1}$ and $V_{\mathrm{RN}_1}$ for $\alpha\in \{0.1,0.3,0.5,0.7,0.9\}$.}
  \label{Delomega1}
\end{table}

\begin{table}[H]
  \centering
  \renewcommand{\arraystretch}{1.2}
  \textbf{QNM difference $\bigl(\Delta\omega\bigr)$ for $\{V_{\mathrm{GD}_2},V_{\mathrm{RN}_2}\}$}\\[10pt]
  \fontsize{12}{15}\selectfont
  \begin{adjustbox}{width=\textwidth}
    \setlength{\tabcolsep}{5pt}
    \begin{tabular}{||c||c||
      cc|cc|cc|cc|cc||}
      \hline\hline
      \multirow{2}{*}{$n$} & \multirow{2}{*}{$n_0$}
        & \multicolumn{2}{c|}{$\alpha=0.1$}
        & \multicolumn{2}{c|}{$\alpha=0.3$}
        & \multicolumn{2}{c|}{$\alpha=0.5$}
        & \multicolumn{2}{c|}{$\alpha=0.7$}
        & \multicolumn{2}{c||}{$\alpha=0.9$} \\
      \cline{3-12}
        &  & $\Delta\Re(\omega)$ & $\Delta\Im(\omega)$
        & $\Delta\Re(\omega)$ & $\Delta\Im(\omega)$
        & $\Delta\Re(\omega)$ & $\Delta\Im(\omega)$
        & $\Delta\Re(\omega)$ & $\Delta\Im(\omega)$
        & $\Delta\Re(\omega)$ & $\Delta\Im(\omega)$ \\
      \thickhline
      \multirow{3}{*}{2}
        & 0 & $5.17\times10^{-4}$ & $1.49\times10^{-5}$
            & $1.08\times10^{-5}$ & $4.38\times10^{-4}$
            & $1.91\times10^{-4}$ & $2.13\times10^{-4}$
            & $1.32\times10^{-3}$ & $1.39\times10^{-3}$
            & $2.42\times10^{-4}$ & $1.66\times10^{-3}$ \\
        & 1 & $4.26\times10^{-3}$ & $2.17\times10^{-3}$
            & $1.58\times10^{-4}$ & $1.32\times10^{-3}$
            & $3.13\times10^{-3}$ & $8.76\times10^{-4}$
            & $4.68\times10^{-4}$ & $3.80\times10^{-3}$
            & $2.88\times10^{-3}$ & $3.59\times10^{-3}$ \\
        & 2 & $1.75\times10^{-2}$ & $1.97\times10^{-2}$
            & $4.72\times10^{-4}$ & $2.24\times10^{-3}$
            & $8.72\times10^{-3}$ & $4.82\times10^{-3}$
            & $4.16\times10^{-3}$ & $7.12\times10^{-3}$
            & $1.28\times10^{-2}$ & $3.81\times10^{-3}$ \\
      \hline
      \multirow{4}{*}{3}
        & 0 & $2.98\times10^{-4}$ & $3.03\times10^{-5}$
            & $5.32\times10^{-5}$ & $4.02\times10^{-4}$
            & $4.57\times10^{-4}$ & $5.76\times10^{-4}$
            & $2.98\times10^{-3}$ & $1.34\times10^{-3}$
            & $4.66\times10^{-4}$ & $1.60\times10^{-3}$ \\
        & 1 & $9.07\times10^{-4}$ & $1.14\times10^{-4}$
            & $7.65\times10^{-5}$ & $1.26\times10^{-3}$
            & $6.66\times10^{-4}$ & $2.18\times10^{-3}$
            & $2.27\times10^{-3}$ & $4.11\times10^{-3}$
            & $1.35\times10^{-3}$ & $4.37\times10^{-3}$ \\
        & 2 & $3.85\times10^{-3}$ & $1.98\times10^{-3}$
            & $2.75\times10^{-4}$ & $2.52\times10^{-3}$
            & $3.42\times10^{-3}$ & $2.33\times10^{-3}$
            & $8.49\times10^{-4}$ & $7.15\times10^{-3}$
            & $8.03\times10^{-3}$ & $3.91\times10^{-3}$ \\
        & 3 & $1.26\times10^{-2}$ & $1.11\times10^{-2}$
            & $1.88\times10^{-3}$ & $5.46\times10^{-3}$
            & $1.08\times10^{-2}$ & $2.60\times10^{-3}$
            & $1.27\times10^{-3}$ & $1.07\times10^{-2}$
            & $2.55\times10^{-2}$ & $9.97\times10^{-3}$ \\
      \hline
      \multirow{5}{*}{4}
        & 0 & $3.35\times10^{-4}$ & $3.24\times10^{-5}$
            & $1.29\times10^{-4}$ & $3.90\times10^{-4}$
            & $1.10\times10^{-4}$ & $7.67\times10^{-4}$
            & $4.38\times10^{-3}$ & $1.36\times10^{-3}$
            & $6.47\times10^{-4}$ & $1.58\times10^{-3}$ \\
        & 1 & $2.63\times10^{-4}$ & $1.21\times10^{-4}$
            & $2.52\times10^{-4}$ & $1.17\times10^{-3}$
            & $8.98\times10^{-5}$ & $2.32\times10^{-3}$
            & $3.87\times10^{-3}$ & $4.12\times10^{-3}$
            & $4.64\times10^{-5}$ & $4.72\times10^{-3}$ \\
        & 2 & $1.92\times10^{-4}$ & $4.25\times10^{-4}$
            & $4.44\times10^{-4}$ & $2.00\times10^{-3}$
            & $2.22\times10^{-4}$ & $4.04\times10^{-3}$
            & $2.87\times10^{-3}$ & $7.01\times10^{-3}$
            & $1.00\times10^{-3}$ & $7.83\times10^{-3}$ \\
        & 3 & $1.84\times10^{-3}$ & $1.72\times10^{-3}$
            & $5.49\times10^{-4}$ & $3.03\times10^{-3}$
            & $4.51\times10^{-4}$ & $6.59\times10^{-3}$
            & $1.41\times10^{-3}$ & $1.01\times10^{-2}$
            & $2.05\times10^{-3}$ & $1.12\times10^{-2}$ \\
        & 4 & $6.06\times10^{-3}$ & $6.05\times10^{-3}$
            & $2.84\times10^{-4}$ & $4.69\times10^{-3}$
            & $3.27\times10^{-3}$ & $1.18\times10^{-2}$
            & $4.70\times10^{-4}$ & $1.35\times10^{-2}$
            & $2.28\times10^{-3}$ & $1.58\times10^{-2}$ \\
      \hline
      \multirow{6}{*}{5}
        & 0 & $4.34\times10^{-4}$ & $2.98\times10^{-5}$
            & $1.84\times10^{-4}$ & $3.85\times10^{-4}$
            & $3.12\times10^{-4}$ & $2.05\times10^{-4}$
            & $5.69\times10^{-3}$ & $1.37\times10^{-3}$
            & $7.37\times10^{-4}$ & $1.57\times10^{-3}$ \\
        & 1 & $4.34\times10^{-4}$ & $9.30\times10^{-5}$
            & $2.59\times10^{-4}$ & $1.16\times10^{-3}$
            & $1.56\times10^{-4}$ & $2.27\times10^{-3}$
            & $5.28\times10^{-3}$ & $4.13\times10^{-3}$
            & $1.81\times10^{-4}$ & $4.69\times10^{-3}$ \\
        & 2 & $3.72\times10^{-4}$ & $1.90\times10^{-4}$
            & $2.59\times10^{-4}$ & $2.02\times10^{-3}$
            & $6.86\times10^{-4}$ & $3.75\times10^{-3}$
            & $4.48\times10^{-3}$ & $6.97\times10^{-3}$
            & $9.99\times10^{-4}$ & $7.68\times10^{-3}$ \\
        & 3 & $6.85\times10^{-5}$ & $4.49\times10^{-4}$
            & $2.58\times10^{-4}$ & $3.28\times10^{-3}$
            & $1.59\times10^{-3}$ & $5.12\times10^{-3}$
            & $3.31\times10^{-3}$ & $9.93\times10^{-3}$
            & $2.97\times10^{-3}$ & $1.04\times10^{-2}$ \\
        & 4 & $8.25\times10^{-4}$ & $1.24\times10^{-3}$
            & $2.17\times10^{-3}$ & $5.83\times10^{-3}$
            & $1.88\times10^{-3}$ & $3.77\times10^{-3}$
            & $1.78\times10^{-3}$ & $1.30\times10^{-2}$
            & $6.00\times10^{-3}$ & $1.23\times10^{-2}$ \\
        & 5 & $2.83\times10^{-3}$ & $3.35\times10^{-3}$
            & $6.81\times10^{-3}$ & $1.17\times10^{-2}$
            & $5.74\times10^{-3}$ & $8.28\times10^{-3}$
            & $6.55\times10^{-5}$ & $1.63\times10^{-2}$
            & $1.05\times10^{-2}$ & $1.28\times10^{-2}$ \\
      \hline
      \multirow{7}{*}{6}
        & 0 & $5.25\times10^{-4}$ & $2.86\times10^{-5}$
            & $2.40\times10^{-4}$ & $3.82\times10^{-4}$
            & $3.87\times10^{-4}$ & $2.03\times10^{-4}$
            & $6.96\times10^{-3}$ & $1.38\times10^{-3}$
            & $8.40\times10^{-4}$ & $1.57\times10^{-3}$ \\
        & 1 & $5.34\times10^{-4}$ & $8.68\times10^{-5}$
            & $3.39\times10^{-4}$ & $1.15\times10^{-3}$
            & $4.41\times10^{-4}$ & $6.10\times10^{-4}$
            & $6.62\times10^{-3}$ & $4.15\times10^{-3}$
            & $3.88\times10^{-4}$ & $4.68\times10^{-3}$ \\
        & 2 & $5.27\times10^{-4}$ & $1.56\times10^{-4}$
            & $5.33\times10^{-4}$ & $1.91\times10^{-3}$
            & $4.91\times10^{-4}$ & $3.77\times10^{-3}$
            & $5.94\times10^{-3}$ & $6.97\times10^{-3}$
            & $4.92\times10^{-4}$ & $7.75\times10^{-3}$ \\
        & 3 & $4.29\times10^{-4}$ & $2.79\times10^{-4}$
            & $7.98\times10^{-4}$ & $2.69\times10^{-3}$
            & $9.84\times10^{-4}$ & $5.30\times10^{-3}$
            & $4.94\times10^{-3}$ & $9.87\times10^{-3}$
            & $1.71\times10^{-3}$ & $1.08\times10^{-2}$ \\
        & 4 & $8.36\times10^{-5}$ & $5.86\times10^{-4}$
            & $1.07\times10^{-3}$ & $3.54\times10^{-3}$
            & $1.44\times10^{-3}$ & $6.97\times10^{-3}$
            & $3.65\times10^{-3}$ & $1.29\times10^{-2}$
            & $3.04\times10^{-3}$ & $1.38\times10^{-2}$ \\
        & 5 & $7.68\times10^{-4}$ & $1.37\times10^{-3}$
            & $1.25\times10^{-3}$ & $4.55\times10^{-3}$
            & $1.58\times10^{-3}$ & $2.64\times10^{-3}$
            & $2.06\times10^{-3}$ & $1.60\times10^{-2}$
            & $4.10\times10^{-3}$ & $1.73\times10^{-2}$ \\
        & 6 & $2.47\times10^{-3}$ & $3.20\times10^{-3}$
            & $1.18\times10^{-3}$ & $5.97\times10^{-3}$
            & $1.39\times10^{-3}$ & $1.72\times10^{-2}$
            & $4.34\times10^{-3}$ & $2.19\times10^{-2}$
            & $4.04\times10^{-3}$ & $1.92\times10^{-2}$ \\
      \hline\hline
    \end{tabular}
  \end{adjustbox}
  \caption{\footnotesize Real and imaginary parts of the difference $\Delta\omega$ between QNMs of potentials $V_{\mathrm{GD}_2}$ and $V_{\mathrm{RN}_2}$ for $\alpha\in \{0.1,0.3,0.5,0.7,0.9\}$.}
  \label{Delomega2}
\end{table}

\begin{table}[H]
  \centering
  \renewcommand{\arraystretch}{1.2}
  \textbf{QNM difference $\bigl(\Delta\omega\bigr)$ for $\{V_{\mathrm{GD}_3},V_{\mathrm{RN}_3}\}$}\\[10pt]
  \fontsize{12}{15}\selectfont
  \begin{adjustbox}{width=\textwidth}
    \setlength{\tabcolsep}{5pt}
    \begin{tabular}{||c||c||
      cc|cc|cc|cc|cc||}
      \hline\hline
      \multirow{2}{*}{$n$} & \multirow{2}{*}{$n_0$}
        & \multicolumn{2}{c|}{$\alpha=0.1$}
        & \multicolumn{2}{c|}{$\alpha=0.3$}
        & \multicolumn{2}{c|}{$\alpha=0.5$}
        & \multicolumn{2}{c|}{$\alpha=0.7$}
        & \multicolumn{2}{c||}{$\alpha=0.9$} \\
      \cline{3-12}
        &  & $\Delta\Re(\omega)$ & $\Delta\Im(\omega)$
        & $\Delta\Re(\omega)$ & $\Delta\Im(\omega)$
        & $\Delta\Re(\omega)$ & $\Delta\Im(\omega)$
        & $\Delta\Re(\omega)$ & $\Delta\Im(\omega)$
        & $\Delta\Re(\omega)$ & $\Delta\Im(\omega)$ \\
      \thickhline
      \multirow{3}{*}{2}
        & 0 & $1.91\times10^{-4}$ & $1.97\times10^{-4}$
            & $5.71\times10^{-4}$ & $5.92\times10^{-4}$
            & $9.48\times10^{-4}$ & $9.89\times10^{-4}$
            & $1.32\times10^{-3}$ & $1.39\times10^{-3}$
            & $1.69\times10^{-3}$ & $1.79\times10^{-3}$ \\
        & 1 & $5.09\times10^{-4}$ & $2.50\times10^{-3}$
            & $2.57\times10^{-4}$ & $1.34\times10^{-3}$
            & $3.56\times10^{-4}$ & $2.57\times10^{-3}$
            & $4.68\times10^{-4}$ & $3.80\times10^{-3}$
            & $5.93\times10^{-4}$ & $5.05\times10^{-3}$ \\
        & 2 & $8.88\times10^{-4}$ & $4.64\times10^{-3}$
            & $1.67\times10^{-3}$ & $2.46\times10^{-3}$
            & $2.90\times10^{-3}$ & $4.76\times10^{-3}$
            & $4.16\times10^{-3}$ & $7.12\times10^{-3}$
            & $5.46\times10^{-3}$ & $9.54\times10^{-3}$ \\
      \hline
      \multirow{4}{*}{3}
        & 0 & $4.26\times10^{-4}$ & $1.91\times10^{-4}$
            & $1.28\times10^{-3}$ & $5.74\times10^{-4}$
            & $2.13\times10^{-3}$ & $9.58\times10^{-4}$
            & $2.98\times10^{-3}$ & $1.34\times10^{-3}$
            & $3.83\times10^{-3}$ & $1.73\times10^{-3}$ \\
        & 1 & $3.26\times10^{-4}$ & $5.85\times10^{-4}$
            & $9.77\times10^{-4}$ & $1.76\times10^{-3}$
            & $1.62\times10^{-3}$ & $2.93\times10^{-3}$
            & $2.27\times10^{-3}$ & $4.11\times10^{-3}$
            & $2.91\times10^{-3}$ & $5.29\times10^{-3}$ \\
        & 2 & $1.28\times10^{-4}$ & $1.02\times10^{-3}$
            & $3.77\times10^{-4}$ & $3.06\times10^{-3}$
            & $6.17\times10^{-4}$ & $5.10\times10^{-3}$
            & $8.49\times10^{-4}$ & $7.15\times10^{-3}$
            & $1.07\times10^{-3}$ & $9.20\times10^{-3}$ \\
        & 3 & $1.68\times10^{-4}$ & $1.53\times10^{-3}$
            & $5.16\times10^{-4}$ & $4.58\times10^{-3}$
            & $8.84\times10^{-4}$ & $7.64\times10^{-3}$
            & $1.27\times10^{-3}$ & $1.07\times10^{-2}$
            & $1.67\times10^{-3}$ & $1.38\times10^{-2}$ \\
      \hline
      \multirow{5}{*}{4}
        & 0 & $6.26\times10^{-4}$ & $1.93\times10^{-4}$
            & $1.88\times10^{-3}$ & $5.80\times10^{-4}$
            & $3.13\times10^{-3}$ & $9.68\times10^{-4}$
            & $4.38\times10^{-3}$ & $1.36\times10^{-3}$
            & $5.63\times10^{-3}$ & $1.75\times10^{-3}$ \\
        & 1 & $5.55\times10^{-4}$ & $5.86\times10^{-4}$
            & $1.66\times10^{-3}$ & $1.76\times10^{-3}$
            & $2.77\times10^{-3}$ & $2.94\times10^{-3}$
            & $3.87\times10^{-3}$ & $4.12\times10^{-3}$
            & $4.97\times10^{-3}$ & $5.30\times10^{-3}$ \\
        & 2 & $4.15\times10^{-4}$ & $9.97\times10^{-4}$
            & $1.24\times10^{-3}$ & $3.00\times10^{-3}$
            & $2.06\times10^{-3}$ & $5.00\times10^{-3}$
            & $2.87\times10^{-3}$ & $7.01\times10^{-3}$
            & $3.68\times10^{-3}$ & $9.02\times10^{-3}$ \\
        & 3 & $2.12\times10^{-4}$ & $1.44\times10^{-3}$
            & $6.25\times10^{-4}$ & $4.32\times10^{-3}$
            & $1.03\times10^{-3}$ & $7.21\times10^{-3}$
            & $1.41\times10^{-3}$ & $1.01\times10^{-2}$
            & $1.79\times10^{-3}$ & $1.30\times10^{-2}$ \\
        & 4 & $4.90\times10^{-5}$ & $1.92\times10^{-3}$
            & $1.65\times10^{-4}$ & $5.77\times10^{-3}$
            & $3.06\times10^{-4}$ & $9.63\times10^{-3}$
            & $4.70\times10^{-4}$ & $1.35\times10^{-2}$
            & $6.59\times10^{-4}$ & $1.73\times10^{-2}$ \\
      \hline
      \multirow{6}{*}{5}
        & 0 & $8.13\times10^{-4}$ & $1.95\times10^{-4}$
            & $2.44\times10^{-3}$ & $5.85\times10^{-4}$
            & $4.06\times10^{-3}$ & $9.76\times10^{-4}$
            & $5.69\times10^{-3}$ & $1.37\times10^{-3}$
            & $7.32\times10^{-3}$ & $1.76\times10^{-3}$ \\
        & 1 & $7.56\times10^{-4}$ & $5.88\times10^{-4}$
            & $2.27\times10^{-3}$ & $1.77\times10^{-3}$
            & $3.78\times10^{-3}$ & $2.95\times10^{-3}$
            & $5.28\times10^{-3}$ & $4.13\times10^{-3}$
            & $6.79\times10^{-3}$ & $5.32\times10^{-3}$ \\
        & 2 & $6.44\times10^{-4}$ & $9.92\times10^{-4}$
            & $1.93\times10^{-3}$ & $2.98\times10^{-3}$
            & $3.21\times10^{-3}$ & $4.97\times10^{-3}$
            & $4.48\times10^{-3}$ & $6.97\times10^{-3}$
            & $5.75\times10^{-3}$ & $8.98\times10^{-3}$ \\
        & 3 & $4.80\times10^{-4}$ & $1.41\times10^{-3}$
            & $1.43\times10^{-3}$ & $4.25\times10^{-3}$
            & $2.38\times10^{-3}$ & $7.09\times10^{-3}$
            & $3.31\times10^{-3}$ & $9.93\times10^{-3}$
            & $4.23\times10^{-3}$ & $1.28\times10^{-2}$ \\
        & 4 & $2.69\times10^{-4}$ & $1.86\times10^{-3}$
            & $7.93\times10^{-4}$ & $5.58\times10^{-3}$
            & $1.30\times10^{-3}$ & $9.31\times10^{-3}$
            & $1.78\times10^{-3}$ & $1.30\times10^{-2}$
            & $2.25\times10^{-3}$ & $1.68\times10^{-2}$ \\
        & 5 & $1.44\times10^{-5}$ & $2.33\times10^{-3}$
            & $1.89\times10^{-5}$ & $6.99\times10^{-3}$
            & $5.69\times10^{-6}$ & $1.16\times10^{-2}$
            & $6.55\times10^{-5}$ & $1.63\times10^{-2}$
            & $1.54\times10^{-4}$ & $2.10\times10^{-2}$ \\
      \hline
      \multirow{7}{*}{6}
        & 0 & $9.93\times10^{-4}$ & $1.96\times10^{-4}$
            & $2.98\times10^{-3}$ & $5.88\times10^{-4}$
            & $4.97\times10^{-3}$ & $9.81\times10^{-4}$
            & $6.96\times10^{-3}$ & $1.38\times10^{-3}$
            & $8.94\times10^{-3}$ & $1.77\times10^{-3}$ \\
        & 1 & $9.46\times10^{-4}$ & $5.89\times10^{-4}$
            & $2.84\times10^{-3}$ & $1.77\times10^{-3}$
            & $4.73\times10^{-3}$ & $2.96\times10^{-3}$
            & $6.62\times10^{-3}$ & $4.15\times10^{-3}$
            & $8.50\times10^{-3}$ & $5.34\times10^{-3}$ \\
        & 2 & $8.51\times10^{-4}$ & $9.91\times10^{-4}$
            & $2.55\times10^{-3}$ & $2.98\times10^{-3}$
            & $4.25\times10^{-3}$ & $4.97\times10^{-3}$
            & $5.94\times10^{-3}$ & $6.97\times10^{-3}$
            & $7.63\times10^{-3}$ & $8.97\times10^{-3}$ \\
        & 3 & $7.13\times10^{-4}$ & $1.40\times10^{-3}$
            & $2.13\times10^{-3}$ & $4.22\times10^{-3}$
            & $3.54\times10^{-3}$ & $7.04\times10^{-3}$
            & $4.94\times10^{-3}$ & $9.87\times10^{-3}$
            & $6.34\times10^{-3}$ & $1.27\times10^{-2}$ \\
        & 4 & $5.33\times10^{-4}$ & $1.83\times10^{-3}$
            & $1.59\times10^{-3}$ & $5.50\times10^{-3}$
            & $2.62\times10^{-3}$ & $9.18\times10^{-3}$
            & $3.65\times10^{-3}$ & $1.29\times10^{-2}$
            & $4.65\times10^{-3}$ & $1.66\times10^{-2}$ \\
        & 5 & $3.14\times10^{-4}$ & $2.28\times10^{-3}$
            & $9.22\times10^{-4}$ & $6.84\times10^{-3}$
            & $2.48\times10^{-3}$ & $1.87\times10^{-2}$
            & $2.06\times10^{-3}$ & $1.60\times10^{-2}$
            & $2.59\times10^{-3}$ & $2.06\times10^{-2}$ \\
        & 6 & $5.93\times10^{-5}$ & $2.74\times10^{-3}$
            & $1.47\times10^{-4}$ & $8.22\times10^{-3}$
            & $1.06\times10^{-3}$ & $1.78\times10^{-3}$
            & $2.09\times10^{-4}$ & $2.47\times10^{-2}$
            & $1.79\times10^{-4}$ & $2.47\times10^{-2}$ \\
      \hline\hline
    \end{tabular}
  \end{adjustbox}
  \caption{\footnotesize Real and imaginary parts of the difference $\Delta\omega$ between QNMs of potentials $V_{\mathrm{GD}_3}$ and $V_{\mathrm{RN}_3}$ for $\alpha\in \{0.1,0.3,0.5,0.7,0.9\}$.}
  \label{Delomega3}
\end{table}

\bibliography{bib_DSS}


\end{document}